\documentclass[conference]{IEEEtran}
\usepackage{cite}

\usepackage{goldschmidt_ieee}

\usepackage{graphicx}
\usepackage{standalone}
\usepackage{float} 
\usepackage[section]{placeins}

\usepackage[dvipsnames]{xcolor}

\title{Data-Driven Hamiltonian Reduction for Superconducting Qubits via Meta-Learning}

\author{
\IEEEauthorblockN{
Arielle Sanford\IEEEauthorrefmark{2}\IEEEauthorrefmark{1},
Andrew T. Kamen\IEEEauthorrefmark{3}\IEEEauthorrefmark{1},
Frederic T. Chong\IEEEauthorrefmark{2},
Andy J. Goldschmidt\IEEEauthorrefmark{4}
}
\IEEEauthorblockA{\IEEEauthorrefmark{2}Department of Computer Science, University of Chicago, Chicago, IL 60637}
\IEEEauthorrefmark{3}Pritzker School of Molecular Engineering, University of Chicago, Chicago, IL 60637
\IEEEauthorblockA{\IEEEauthorrefmark{4}Johns Hopkins Applied Physics Laboratory, Laurel, MD 20723}
\IEEEauthorblockA{\IEEEauthorrefmark{1}These authors contributed equally. Corresponding author: arielles@uchicago.edu}
}

\begin{document}
\maketitle

\thispagestyle{plain}
\pagestyle{plain}

\begin{abstract}

We introduce HAML (Hamiltonian Adaptation via Meta-Learning), a framework for fast online adaptation of effective Hamiltonian models of superconducting quantum processors. 
HAML proceeds in two phases. 
A supervised training phase uses an ensemble of simulated devices to learn an offline map from control inputs and device parameters to effective Hamiltonian coefficients. 
An online adaptation phase then uses a small number of hardware-accessible measurements to identify the unknown parameters of a new device. 
By training directly against effective two-qubit coefficients extracted from full multi-mode simulations, HAML implicitly learns the reduction from full multi-mode Hamiltonians to effective qubit descriptions without invoking perturbation theory. 
We further show that a variance-maximizing greedy selection of measurement configurations boosts online adaptation efficiency. 
We demonstrate HAML on a transmon-coupler-transmon system, recovering effective two-qubit coefficients across a wide range of operating regimes, including parameter regions where Schrieffer-Wolff perturbation theory (SWPT) breaks down. 
This establishes a scalable, sample-efficient approach to Hamiltonian reduction and characterization for near-term quantum processors, with direct implications for calibration, control, and error mitigation.

\end{abstract}


%
\section{Introduction} \label{sec:introduction}

Modern superconducting quantum processors are physically multi-mode devices: each computational qubit is coupled to auxiliary modes such as tunable couplers~\cite{yan2018} and readout resonators that mediate interactions and measurement. The control and calibration stack that drives these processors, however, operates almost exclusively on an effective qubit-level Hamiltonian. Bridging the gap between the full multi-mode device dynamics and the reduced qubit subspace description is therefore essential for high-fidelity gate design, accurate calibration, and effective error mitigation.

A standard analytical bridge is Schrieffer-Wolff perturbation theory (SWPT)~\cite{bravyi2011swpt}, which constructs a unitary transformation $U = e^S$ that, order by order in a small expansion parameter, approximately block-diagonalizes the Hamiltonian into a low-energy (qubit) subspace and a high-energy subspace that can be discarded. SWPT is widely used and physically transparent~\cite{blais2021circuit}, but it carries two costs. First, its convergence requires the ratio of the qubit-coupler coupling to the qubit-coupler frequency difference to be much less than one, $g_{qc}/\Delta_{qc} \ll 1$ with $\Delta_{qc} = \omega_q - \omega_c$. Two-qubit gate speeds are inversely related to this frequency difference, $\Delta_{qc}$; fast gates straddle an intermediate-detuning regime where a qubit description endures even as the accuracy of SWPT starts to deteriorate. Second, the SWPT symbolic derivation can become unwieldy as the system size increases; moreover, numerical approaches can be expensive if recomputation is needed frequently, like in ``inner loop'' applications such as control optimization~\cite{UnitaryTransformations.jl}. Third, in order to arrive at the SWPT Hamiltonian coefficients, the bare coefficients from the full Hamiltonian are required as inputs--including those of the coupler. However, in both academic and commercial superconducting devices, the coupler elements rarely have dedicated readout resonators~\cite{zhang2023characterization, li2020, mundada2019, wu2024, wu2025, gambetta2016, stehlik2021, arute2019quantum, google2025quantum}.

Data-driven approaches to quantum characterization and control offer an alternative path. Recent work has demonstrated graybox modeling that fuses physics priors with machine learning~\cite{goldschmidt2021bilinear, youssry2024graybox}, machine-learning-based optimal control on superconducting qubits~\cite{genois2025quantum}, and model-free reinforcement-learning policies trained directly on hardware measurements~\cite{sivak2022modelfree, wu2026}. These approaches focus on learning high-performance controllers of particular quantum devices or noise environments, but the learned models or policies are not explicitly designed for rapid transfer or adaptation.

Meta-learning~\cite{finn2017maml, zintgraf2019cavia} is a particularly good match for device characterization and control. The setting it was designed for---many related tasks with only a few examples each---matches two situations that arise naturally in superconducting hardware. First, quantum processors contain many qubits that are nominally similar but fabrication-distinct, each requiring its own characterization and calibration under tight measurement budgets~\cite{vandamme2024advanced}. This is particularly relevant for modular quantum processors, in which multiple qubits and couplers must be calibrated to operate across varying device configurations~\cite{wu2024}. Second, device parameters drift over time due to noise, so the same chip recalibrated at different times effectively presents as a sequence of similar but non-identical devices~\cite{burnett2019decoherence}. Meta-learning has already been applied to learn generalizable dynamics across a family of related physical systems~\cite{li2023imode}. Meta-learning has also been applied directly for few-shot learning of closed and open quantum systems, demonstrating its promise for transferable modeling across diverse quantum dynamical regimes~\cite{schorling2025metalearning}. In our work, we extend this approach by uniting the characterization and control problem in one framework, meta-learning control models that rapidly adapt across entire families of quantum devices. In addition, to the best of our knowledge, we introduce the first example of a data-driven, control-aware reduction of full multi-mode dynamics to an effective qubit Hamiltonian using sim-to-real adaptation in place of analytic derivations.

We call our framework HAML (Hamiltonian Adaptation via Meta-Learning), and emphasize the following contributions: 
(i) a meta-learning framework that learns the map from device parameters and control inputs to effective Hamiltonian coefficients, trained across an ensemble of devices offline and adapted to a new device online from only a handful of measurements; 
(ii) a data-driven model reduction that recovers the projected two-qubit Hamiltonian from full three-mode dynamics without perturbation theory,
%
(iii) a greedy approach for selecting information-seeking (state, observable) measurement pairs and control pulses, which enable sample-efficient adaptation to new devices,
%
and (iv) a demonstration that the learned reduction remains accurate in regimes where leading-order SWPT visibly breaks down.

The manuscript contains the following sections: 
Sec.~\ref{sec:background} introduces the relevant background on machine learning, quantum hardware, and perturbation theory: the meta-learning framework (Sec.~\ref{sub:metalearning}), the transmon-coupler-transmon model Hamiltonian (Sec.~\ref{sub:system}), and the Schrieffer-Wolff transformation (Sec.~\ref{sub:swpt}). Sec.~\ref{sec:methods} presents HAML's two stage framework (Sec.~\ref{sub:formulation}): (1) offline training (Sec~\ref{sub:datagen}) and (2) online adaptation (Sec.~\ref{sub:adaptation}). To minimize the measurement budget during adaptation, information-seeking initial states, observables, and fluxes are chosen (Sec.~\ref{sub:measurement}). Sec.~\ref{sec:results} reports HAML's accuracy against SWPT on 10 held-out test devices: per-coefficient prediction errors (Sec.~\ref{sub:coeff_results}), the range of qubit-coupler hybridization spanned by the test ensemble (Sec.~\ref{sub:hybridization}), and projected-unitary fidelity comparisons across the operating range (Sec.~\ref{sub:infid_results}).
The conclusion (Sec~\ref{sec:conclusion}) follows.

\section{Background} \label{sec:background}
\subsection{Learning Across Device Ensembles} \label{sub:metalearning}

HAML trains a single neural network to represent the dynamics of a quantum device. The network can be understood as a physics-informed, neural ordinary differential equation~\cite{chen2018neural} efficiently representing the Hamiltonian~\cite{genois2025quantum} (see Methods~\ref{sec:methods}). The training data are drawn from an ensemble of distinct instances of parameterized quantum devices. Each training point includes control inputs and parameter values. After training, the network is frozen and a small set of adaptation parameters, $\eta$, is optimized per device using a handful of hardware measurements.

This structure is inspired by the meta-learning literature. Model-Agnostic Meta-Learning (MAML)~\cite{finn2017maml} introduced the idea of training a network whose weights serve as a good initialization for fast gradient-based adaptation on a new task at test time. Fast Context Adaptation via Meta-Learning (CAVIA)~\cite{zintgraf2019cavia} refined this by partitioning parameters into two sets: shared weights, which are meta-trained, and a small set of context parameters, which are adapted per task. CAVIA both reduces meta-overfitting and yields a low-dimensional representation for efficient task adaptation. 
HAML adopts the parameter split of CAVIA but is simpler in one important respect: rather than learning the context inputs via a second, inner-loop optimization during training, $\eta$ is supplied as a known input during the meta-training phase. This is possible because training is performed on simulation, where ground-truth $\eta$ values are always available. In our formulation, the only gradient-based optimization of $\eta$ occurs during online adaptation. Upon deployment, $\eta$ is unknown, and its estimation becomes a system identification problem in the space of context parameters. We also use the learned model to select experiments that improve the adaptation efficiency.

Because context is learned offline and inferred online, our approach can be viewed as a sim-to-real variant of CAVIA-style context adaptation. In quantum engineering, the Hamiltonian operator structure is usually well-specified by device physics, while uncertainty is concentrated within a few device-dependent parameters~\cite{blais2021circuit}. This makes simulation a reliable way to generate realistic instances of device contexts. That being said, identifying the parameters of a full Hamiltonian model can be challenging for a number of reasons like system size or missing experimental access (see Sec.~\ref{sub:swpt}); this motivates the need for effective descriptions that are limited to the accessible degrees of freedom.
Indeed, while HAML is in principle general to any quantum device family, it is especially well matched to systems with auxiliary modes, such as couplers or Purcell filters, because these mediate interactions but cannot be measured directly (e.g., ~\cite{yan2018, wu2024}). In such systems, the network must infer the auxiliary-mode contribution from qubit-accessible measurements alone, a task at which meta-learned priors are particularly adept.

\begin{figure}[t]
    \centering
    \includegraphics[width=\columnwidth]{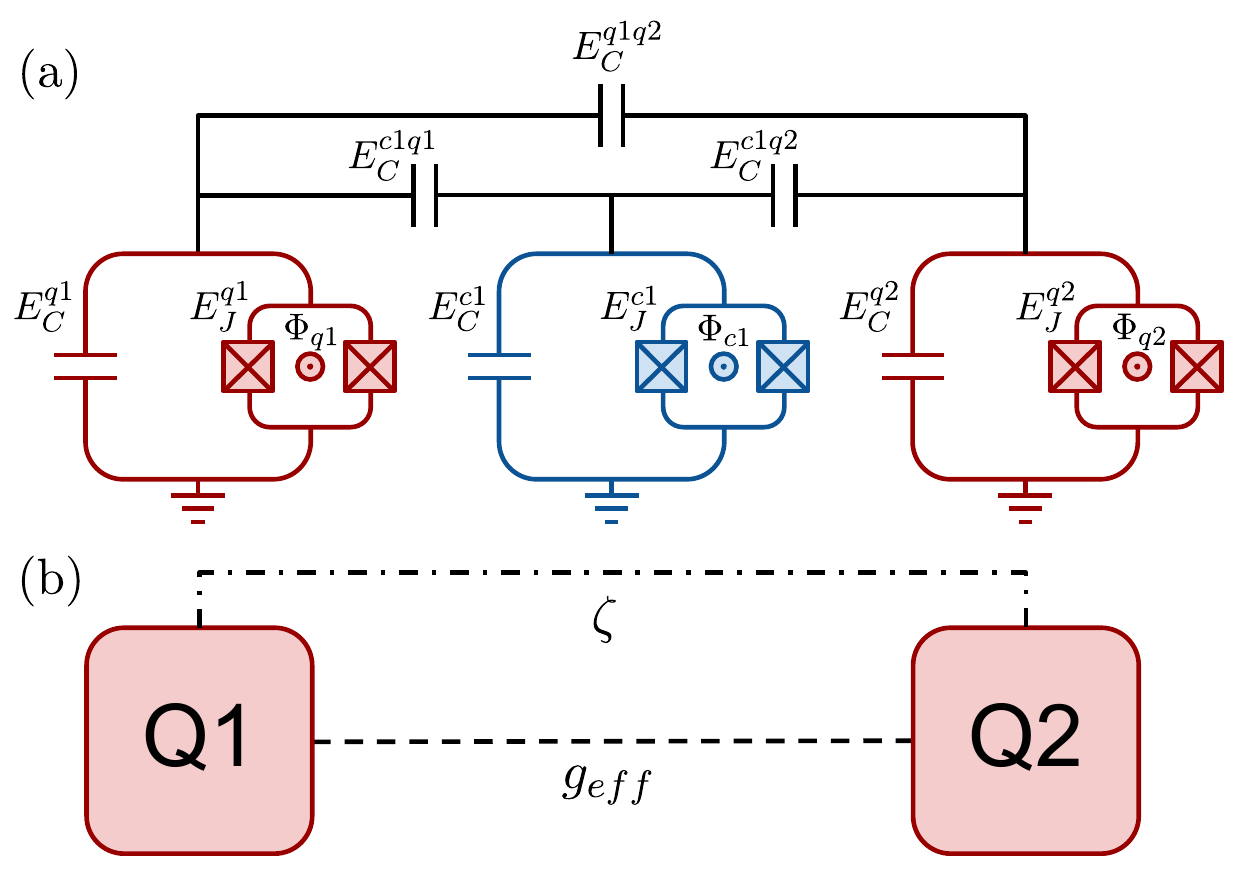}
    \caption{(a) Circuit diagram of the transmon-coupler-transmon architecture: two frequency-tunable transmon qubits (red) capacitively coupled to a tunable coupler (blue). Each mode is characterized by its charging and Josephson energies $E_C^i$, $E_J^i$, with capacitive coupling energies $E_C^{jk}$ setting the inter-mode couplings. The qubit and coupler frequencies are tunable via the SQUID fluxes $\Phi_{q1}$, $\Phi_{q2}$, $\Phi_{c1}$. (b) Effective two-qubit description after coupler elimination: the qubits acquire an effective exchange coupling $g_{\text{eff}}$ and a residual $ZZ$ interaction $\zeta$.}
    \label{fig:background}
\end{figure}
\subsection{The Transmon-Coupler System} \label{sub:system}

In this work, we explore the utility of meta-learning for accurate characterization of a system consisting of two superconducting qubits. Specifically, we consider an architecture (see Fig.~\ref{fig:background}) in which two transmon qubits have a shared coupler that mediates a virtual exchange interaction \cite{yan2018}. To model the dynamics, we use the following Hamiltonian, 
\begin{align}
    \frac{H_{\text{full}}}{\hbar} = \sum_{i\in\mathcal{M}} \frac{\Delta_i}{2} Z_i + \sum_{j,k\in\mathcal{M}} \frac{g_{jk}}{2} (X_j X_k + Y_j Y_k)
    \label{eq:h-full}
\end{align}
in which we have already applied the rotating wave approximation. For simplicity, we have also restricted ourselves to a two-level approximation of the respective modes, $\mathcal{M}:=\{q1, q2, c1\}$, corresponding to the two qubits and the coupler. $\Delta_i = \omega_i -\omega_0^i$ is the detuning of mode $i$ with respect to the rotating frame $\omega_0^i$, and $g_{jk}$ is the coupling rate between mode $j$ and mode $k$. The mode frequencies and coupling rates are tunable via external magnetic flux. More concretely, 
\begin{align}
    \omega_i / 2\pi &= \sqrt{8E_J^i E_C^i}-E_C^i   \\
    g_{j k} / 2\pi &= \frac{E_C^{jk}}{\sqrt{2}}\left(\frac{E_J^j E_J^{k}}{E_C^j E_C^k} \right)^{1/4}
\end{align}
where $E_J=E_{J0}\left| \cos{(\pi \Phi_{\text{ext}}/\Phi_0)} \right|=E_{J0}\left|\cos{(\phi_{\text{ext}}/2)}\right|$, assuming symmetric SQUIDs, with $\Phi_0=h/2e$ denoting the flux quantum, and $\phi_{\text{ext}}$ representing the reduced external flux. Here, $E_J^i$ and $E_C^i$ are the Josephson and charging energies of mode $i$, and $E_C^{j k}$ is the capacitive coupling energy between modes $j$ and $k$.

As noted in Sec.~\ref{sec:introduction}, couplers in superconducting devices typically lack dedicated readout resonators. This design choice reflects the fact that additional readout components introduce hardware overhead, increase control complexity, and contribute to decoherence~\cite{li2025, zhang2025}. Absent dedicated readout for the coupler(s), Hamiltonian characterization of these devices is typically restricted to the qubit subspace(s). The effective qubit-subspace Hamiltonian takes the form,
\begin{align}
    \frac{H_{\text{eff}}}{\hbar} = \sum_{i\in\widetilde{\mathcal{M}}} \frac{\tilde\Delta_i}{2} Z_i + \frac{g_{\text{eff}}}{2} (X_{q1} X_{q2} + Y_{q1} Y_{q2}) +\zeta Z_{q1}Z_{q2}
    \label{eq:h-eff}
\end{align}
which approximates the dominant dynamics for the qubit-only subspace, $\widetilde{\mathcal{M}}:=\{q1, q2\}$, across the dispersive and the intermediate-detuning regimes. $\tilde\Delta_i=\tilde\omega_i - \omega_0^i$ is the difference between the effective qubit frequency and the rotating frame, $g_{\text{eff}}$ is the effective coupling between the qubits, and $\zeta$ is the parasitic $ZZ$ interaction strength. The accuracy of any specific reduction to this form (e.g., Sec.~\ref{sub:swpt}) degrades as the qubit and coupler approach resonance, i.e., as the qubit-coupler detuning $\Delta_{qc}$ approaches zero. In this near-resonance limit, strong hybridization renders any qubit-subspace description ill-defined.

\subsection{The Schrieffer-Wolff Transformation} \label{sub:swpt}
A common analytical approach to obtaining the effective Hamiltonian in Eq.~\eqref{eq:h-eff} is to apply Schrieffer-Wolff perturbation theory (SWPT)~\cite{bravyi2011swpt}, eliminating the qubit-coupler interaction perturbatively via the Schrieffer-Wolff transformation~\cite{yan2018}. To start, we can group the terms in Eq.~\eqref{eq:h-full},
\begin{align}
    H_{\text{full}} = H_q + H_c+ H_{qc}
\end{align}
where $H_q$, $H_c$, $H_{qc}$ correspond to the bare qubit dynamics, the bare coupler dynamics, and the qubit-coupler interaction, respectively. If $|g_{qc}/\Delta_{qc}|\ll 1$, we can apply the unitary ${U}=e^S$ to block-diagonalize the Hamiltonian as follows,
\begin{align}
    e^{S} H e^{-S} = \tilde{H}_q + \tilde{H}_c = H_{q} + H_{c} + \frac{1}{2}[S, H_{qc}] +\mathcal{O}(\epsilon^3)
\end{align}
where $S$ is an anti-Hermitian generator satisfying the condition, 
\begin{align}
    H_{qc}+  [S, H_q + H_c] = 0
\end{align}
and $\epsilon:=g_{qc}/\Delta_{qc}$ is the order parameter in this approximation. For the system described in Eq.~\eqref{eq:h-full}, $S=\sum_{j\in\widetilde{\mathcal{M}}}(g_{jc1}/\Delta_{jc1})(\sigma^+_{j}\sigma_{c1}^- - h.c.)$ gives the second-order approximation for the qubit-subspace Hamiltonian, with an effective coupling,
\begin{align}
    g_{\text{eff}} \approx \frac{g_{q1c1} g_{q2 c1}}{\Delta} + g_{q1q2}
\end{align}
and effective qubit frequency,
\begin{align}
    \tilde\omega_{j} \approx \omega_j + g_{jc1}^2/\Delta_{jc1} 
\end{align}
where $\Delta_{jc1} = \omega_j-\omega_{c1}$ and $1/ \Delta = (1/\Delta_{q1c1} + 1/\Delta_{q2c1})/2$. For simplicity, we take the second-order expressions above as our analytical baseline and benchmark them against our data-driven reduction, which recovers the full effective qubit-subspace Hamiltonian---including $\zeta$---without invoking perturbation theory. A direct comparison against higher-order treatments that recover $\zeta$ analytically~\cite{mundada2019, petterssonfors2024} is left for future work, as it involves significantly more involved calculations.

A few comments on this perturbative approach are in order. First, the convergence of the expansion requires $|\epsilon| = |g_{qc}/\Delta_{qc}| \ll 1$, so the approximation breaks down as the qubit-coupler detuning decreases. This is precisely the regime relevant to fast two-qubit gate operation, where the mediated effective coupling $g_{\text{eff}}$ grows as $\Delta_{qc}$ shrinks. Second, even for the single-coupler architecture treated above, a complete analytical characterization of $\zeta$ requires a significantly more involved derivation~\cite{mundada2019, petterssonfors2024}, and extensions of SWPT to architectures with additional couplers or auxiliary modes require new per-architecture derivations whose complexity grows rapidly with the number of modes. Third, any perturbative reduction takes the bare mode frequencies and couplings as inputs, but these parameters are not directly accessible on hardware architectures where the coupler lacks dedicated readout. Back-calculating them requires indirect probes such as qubit-coupler spectroscopy (e.g. Lamb-shift measurements in Ref.~\cite{wu2024}), adding a nontrivial calibration step that must precede any use of the perturbative reduction.

Together, these considerations motivate a data-driven approach that bypasses the perturbative expansion, generalizes across architectures, and operates directly on qubit-subspace observables. Such an approach has the potential to match or exceed second-order SWPT accuracy, particularly in the intermediate-detuning regime where the perturbative expansion breaks down. We develop this approach in the section that follows.

\section{Methods} \label{sec:methods}
%
\begin{figure*}[th]
    \centering
    \includegraphics[width=0.85\textwidth]{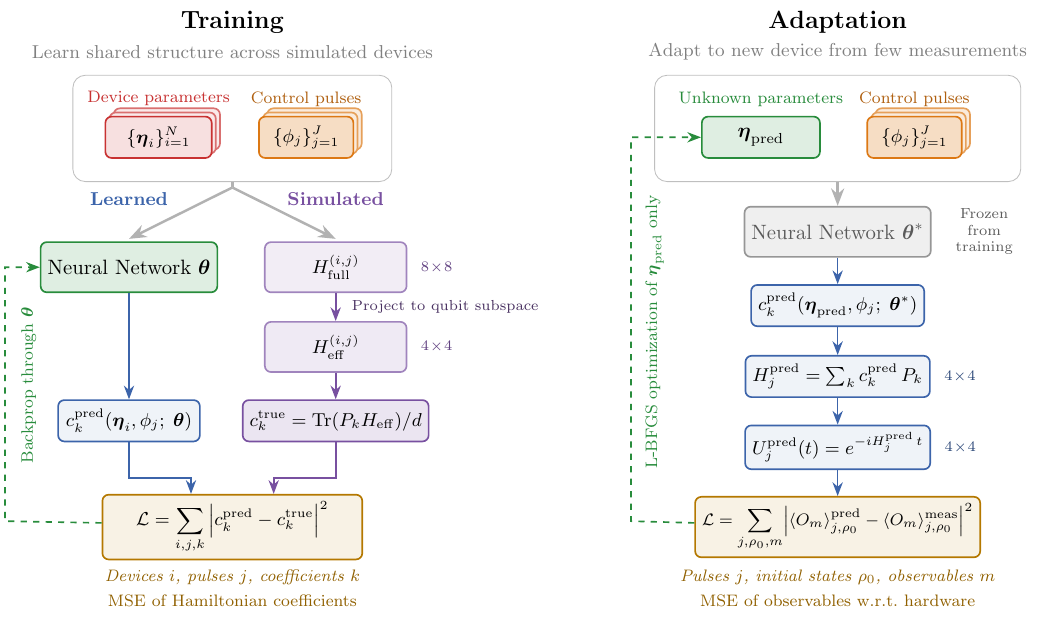}
    \caption{HAML pipeline: schematic of the two-stage framework. Left: during training, a neural network with shared parameters $\boldsymbol{\theta}$ learns to predict effective Hamiltonian coefficients from simulated device parameters and control pulses. Right: during adaptation, the trained network is frozen, and only the device parameter vector $\boldsymbol{\eta}_{\mathrm{pred}}$ is optimized using hardware measurements.}
    \label{fig:pipeline}
\end{figure*}

The full HAML pipeline, supervised training across an ensemble of simulated devices followed by online adaptation to a new device, is summarized in Fig.~\ref{fig:pipeline}.

\subsection{Problem Formulation} \label{sub:formulation}
We address the problem of learning the coefficients of a Hamiltonian with a known Pauli structure as a function of control inputs and unknown device parameters. Let $\{P_k\}$ be a Pauli basis spanning the operators that appear in the relevant subspace. We represent the device Hamiltonian as
\begin{align}
    H(\phi, \eta; \theta) = \sum_k c_k(\phi, \eta; \theta) \, P_k,
    \label{eq:pauli-decomp}
\end{align}
where $\phi$ collects the control inputs, $\eta$ collects unknown device-specific parameters, and $\theta$ are the parameters of a neural network that maps $(\phi, \eta)$ to each coefficient $c_k$. Once $\theta$ is fixed by training, characterizing a new device reduces to determining its $\eta$ from a small number of measurements.

We apply this framework for the transmon-coupler-transmon system of Sec.~\ref{sub:system}. The control input is the external fluxes
\begin{align}
    \phi = (\phi_{ext}^{q1},\, \phi_{ext}^{q2},\, \phi_{ext}^{c1}),
\end{align}
which set the mode frequencies and exchange couplings through the Josephson-energy dependence given in Sec.~\ref{sub:system}. The qubit parameters are taken as known, and the unknown coupler-related parameters form the vector
\begin{align}
    \eta = \{E_{J0}^{c1},\, E_C^{c1},\, E_C^{q1,c1},\, E_C^{q2,c1},\, E_C^{q1,q2}\},
\end{align}
comprising the coupler's bare Josephson and charging energies together with the three capacitive coupling energies between the modes. The Pauli basis $\{P_k\}$ is taken to span the operators that appear in the effective two-qubit Hamiltonian of Sec.~\ref{sub:system}, so that the coefficients $c_k$ to be learned are $\tilde{\Delta}_{q1}$, $\tilde{\Delta}_{q2}$, $g_{\text{eff}}$, and $\zeta$. As discussed in Sec.~\ref{sub:metalearning}, $\eta$ varies from one device to the next due to fabrication variations in a modular processor and to drift over time on any single device, motivating the ensemble-based approach of HAML.

\subsection{Data Generation: Hamiltonian Simulation and Effective Coefficient Extraction} \label{sub:datagen}
Training requires a dataset of $(\phi, \eta)$ inputs paired with ground-truth Pauli coefficients $c^{\text{true}}$ for an ensemble of simulated devices, corresponding to the red and orange input stacks at the top-left of Fig.~\ref{fig:pipeline}. Device parameter vectors $\{\eta_i\}_{i=1}^{N}$ are sampled over physically reasonable ranges meant to span the variability expected across devices in a modular processor and across drift cycles of a single device. For each device, control vectors are drawn uniformly at random from a pulse library $\{\phi_j\}_{j=1}^{J}$ spanning the operating range of the coupler and deliberately including near-resonance regions. Specific sampling distributions and ranges are reported in Appendix~\ref{appendix:hyperparams}.

For each $(\phi_j, \eta_i)$ pair the full three-mode Hamiltonian $H_{\text{full}}^{(i,j)}$ of Sec.~\ref{sub:system} is constructed and reduced to the effective two-qubit Hamiltonian $H_{\text{eff}}^{(i,j)}$ in two steps; this is the purple ``Simulated'' branch of Fig.~\ref{fig:pipeline}. First, a dressed-state projection (next paragraph) yields a spectral Hamiltonian $H_{\text{dress}}^{(i,j)}$ whose Pauli coefficients
\begin{align}
    c_k^{\text{dress}}(\phi_j, \eta_i) = \frac{\operatorname{Tr}\!\bigl(P_k\, H_{\text{dress}}^{(i,j)}\bigr)}{d}
    \label{eq:cLow}
\end{align}
with $d=4$ reproduce the qubit-like spectrum of $H_{\text{full}}^{(i,j)}$. Second, these coefficients are refined to maximize process fidelity to the projected dynamics at the final time $T$, yielding the supervised target $c^{\text{true}}(\phi_j, \eta_i)$ and the corresponding effective Hamiltonian $H_{\text{eff}}^{(i,j)} = \sum_k c_k^{\text{true}}(\phi_j, \eta_i)\, P_k$, defined in Eq.~\ref{eq:c_true_fidelity} below. The same procedure produces ground-truth coefficients for evaluation on held-out devices.

To arrive at the dressed-state Hamiltonian $H_{\text{dress}}$ in the qubit-subspace, we utilize a dressed-state projection inspired by symmetric-orthogonalization techniques from quantum chemistry~\cite{lowdin1950non}. To start, we diagonalize the full Hamiltonian, $H_{\text{full}}\in \mathbb{C}^{8\times 8}$, 
\begin{align}
    H_{\text{full}}\ket{\psi_k} = E_k \ket{\psi_k},\quad k\in\{1,...,8\}
\end{align}
yielding the eigenenergies $E_k$ and the corresponding eigenvectors $\ket{\psi_k}$. Next, to project to the qubit subspace, we identify the eigenvectors with the greatest ``qubit character," which we define in terms of the weights
\begin{align}
    w_k = \sum_{\ell\in \mathcal{Q}} |\bra{\ell}\ket{\psi_k}|^2
\end{align}
where $\mathcal{Q}:=\{\ket{0_{q1} 0_{q2} 0_{c1} }, \ket{0_{q1} 1_{q2} 0_{c1} }, \ket{1_{q1} 0_{q2} 0_{c1} }, \ket{1_{q1} 1_{q2} 0_{c1} }\}$ is the qubit subspace when the coupler is in the ground state. The four eigenstates with the largest weight $w_k$ are selected. We label these $|\psi_\alpha\rangle$ with $\alpha = 1, \dots, 4$, sorted by ascending energy.

In the dispersive regime, each selected eigenstate is predominantly a qubit-subspace state dressed by a small coupler admixture. This selection fails only when the coupler is near-resonant with a qubit, causing strong hybridization. However, this requirement---that the qubit-like and coupler-like eigenstates remain distinguishable by their subspace weight---is considerably weaker than the $g_{qc}/\Delta_{qc} \ll 1$ convergence requirement of perturbative methods such as the second-order Schrieffer--Wolff transformation, allowing the dressed-state projection to remain accurate over a wider range of coupler flux bias points.

After selecting the four ``qubit-like" eigenstates, we define the qubit-subspace overlap matrix as follows, 
\begin{align}
    M_{\ell,\alpha} = \bra{\ell}\ket{\psi_\alpha}, \quad \ell\in\mathcal{Q},\;\alpha\in\{1,...,4\}.
\end{align}
Lastly, we apply a symmetric orthogonalization to the overlap matrix $M$. Specifically, we compute the Gram matrix,
\begin{align}
    S = M^\dag M.
\end{align}
For which we have the eigendecomposition,
\begin{align}
    S = \sum_{j=1}^4\lambda_j \ket{v_j}\bra{v_j}\implies S^{-1/2} = \sum_{j=1}^4 \lambda_j^{-1/2} \ket{v_j}\bra{v_j}.
\end{align}
Then, by symmetric orthogonalization, we have
\begin{align}
    \tilde{M} = M S^{-1/2},
\end{align}
and arrive at the dressed-state Hamiltonian,
\begin{align}
    H_{\text{dress}} &= \tilde{M} \text{diag}(E_1,...,E_4)\tilde{M}^\dag\\&= \sum_{\alpha=1}^4 E_\alpha \ket{\tilde{\psi}_\alpha}\bra{\tilde{\psi}_\alpha}.
\end{align}
where $\ket{\tilde{\psi}_\alpha}$ denotes the $\alpha$-th column in $\tilde{M}$.

The dressed-state coefficients $c^{\text{dress}}$ from Eq.~\eqref{eq:cLow} reproduce the qubit-like spectrum of $H_{\text{full}}$, but the unitary they generate need not coincide with the projected dynamics actually produced by the full three-mode evolution. Define the projected sub-unitary
\begin{align}
    U_{\text{proj}}^{(i,j)}(T) = \Pi_\mathcal{Q}\, e^{-i H_{\text{full}}^{(i,j)} T}\, \Pi_\mathcal{Q},
\end{align}
where $\Pi_\mathcal{Q} = \sum_{\ell \in \mathcal{Q}} |\ell\rangle\!\langle\ell|$ is the projector onto the bare coupler-ground qubit subspace $\mathcal{Q}$ defined above; equivalently, $U_{\text{proj}}(T)$ is the $4 \times 4$ block of the full $8 \times 8$ propagator with both indices restricted to $\mathcal{Q}$. The propagator $\exp(-i \sum_k c_k^{\text{dress}} P_k\, T)$ generated by the spectral coefficients does not in general equal $U_{\text{proj}}^{(i,j)}(T)$, since the latter is not unitary (some amplitude leaks out of $\mathcal{Q}$) and so is not reachable from any two-qubit Hamiltonian. We therefore refine the supervised target by maximizing process fidelity to $U_{\text{proj}}$ at the final time:
\begin{align}
    c^{\text{true}}(\phi_j, \eta_i) = \operatorname*{arg\,max}_{c \in \mathbb{R}^{n}}\;
    \frac{\bigl|\operatorname{Tr}\bigl( e^{i \sum_k c_k P_k T}\, U_{\text{proj}}^{(i,j)}(T) \bigr)\bigr|^2}{d^2},
    \label{eq:c_true_fidelity}
\end{align}
solved with L-BFGS initialized at $c^{\text{dress}}(\phi_j, \eta_i)$. The corresponding effective two-qubit Hamiltonian is $H_{\text{eff}}^{(i,j)} = \sum_k c_k^{\text{true}}(\phi_j, \eta_i)\, P_k$. The seeding is essential: the fidelity objective has $\pi$-spaced local maxima along the $Z$-like coordinates, and an unseeded initialization lands on different branches across $(\phi, \eta)$ samples, producing a multi-valued target the network cannot fit as a smooth function. Seeding at $c^{\text{dress}}$ pins every sample to the basin containing the spectral coefficients, yielding a continuous map $(\phi, \eta) \mapsto c^{\text{true}}$ and therefore an effective Hamiltonian that is the best two-qubit representation of the projected dynamics at $T$.

\subsection{Training} \label{sub:training}
Training proceeds as drawn in the left half of Fig.~\ref{fig:pipeline}. The network $f_\theta$ is a multilayer perceptron with SiLU activations that maps each $(\phi, \eta)$ pair to predicted Pauli coefficients $c_k^{\text{pred}}(\phi, \eta; \theta)$. These predictions are compared to the ground-truth coefficients $c^{\text{true}}$ produced by the procedure of Sec.~\ref{sub:datagen} via the per-coefficient mean-squared error
\begin{align}
    \mathcal{L}_{\text{train}}(\theta) = \sum_{i, j, k} \bigl| c_k^{\text{pred}}(\phi_j, \eta_i; \theta) - c_k^{\text{true}}(\phi_j, \eta_i) \bigr|^2,
\end{align}
indexed over simulated devices $i$, control pulses $j$, and Pauli coefficients $k$. The weights, $\theta$, are updated by standard backpropagation.

The network is trained with Adam over a fixed number of epochs. The architecture and training hyperparameters are reported in Appendix~\ref{appendix:hyperparams}, and convergence behavior is shown in Appendix~\ref{appendix:convergence}.

\subsection{Adaptation to New Devices} \label{sub:adaptation}
Adaptation proceeds along the right half of Fig.~\ref{fig:pipeline}. The network weights $\theta^*$ are frozen and the device parameter vector becomes the only optimization variable, denoted $\eta_{\text{pred}}$ to distinguish it from the ground-truth values of $\eta$ supplied during training. A small set of measurements is performed on the new device in the form of (initial state, observable) pairs $\{(\rho_p, O_p)\}_{p=1}^{N_M}$, evaluated at a fixed final evolution time $T$ across selected control points $\phi$.

For a candidate $\eta_{\text{pred}}$, the network output is mapped to a predicted Hamiltonian $H_j^{\text{pred}} = \sum_k c_k^{\text{pred}}(\eta_{\text{pred}}, \phi_j; \theta^*)\, P_k$ and propagator $U_j^{\text{pred}}(T) = e^{-i H_j^{\text{pred}} T}$. Predicted expectation values $\langle O_p \rangle^{\text{pred}}$ are then compared to the corresponding hardware measurements $\langle O_p \rangle^{\text{meas}}$ via the sum of squared residuals
\begin{align}
    \mathcal{L}_{\text{adapt}}(\eta_{\text{pred}}) = \sum_{j, p} \bigl| \langle O_p \rangle_{j}^{\text{pred}}(\eta_{\text{pred}}) - \langle O_p \rangle_{j}^{\text{meas}} \bigr|^2.
\end{align}
We minimize $\mathcal{L}_{\text{adapt}}$ with L-BFGS over $\eta_{\text{pred}}$ alone, with 5 random initializations to mitigate the risk of poor local minima.

The recovered $\eta_{\text{pred}}$ is not expected to exactly coincide with the device's true parameters: the map $\eta \mapsto c_k$ is not injective, so distinct $\eta$ can produce identical observable physics. We therefore evaluate adaptation by the accuracy of the recovered effective coefficients rather than by recovery of $\eta_{\text{pred}}$ itself. The choice of which $(\rho_p, O_p)$ pairs to use is the subject of the next subsection.

\subsection{Minimal Measurement Set} \label{sub:measurement}
\begin{figure}[t]
    \centering
    \includegraphics[width=\columnwidth]{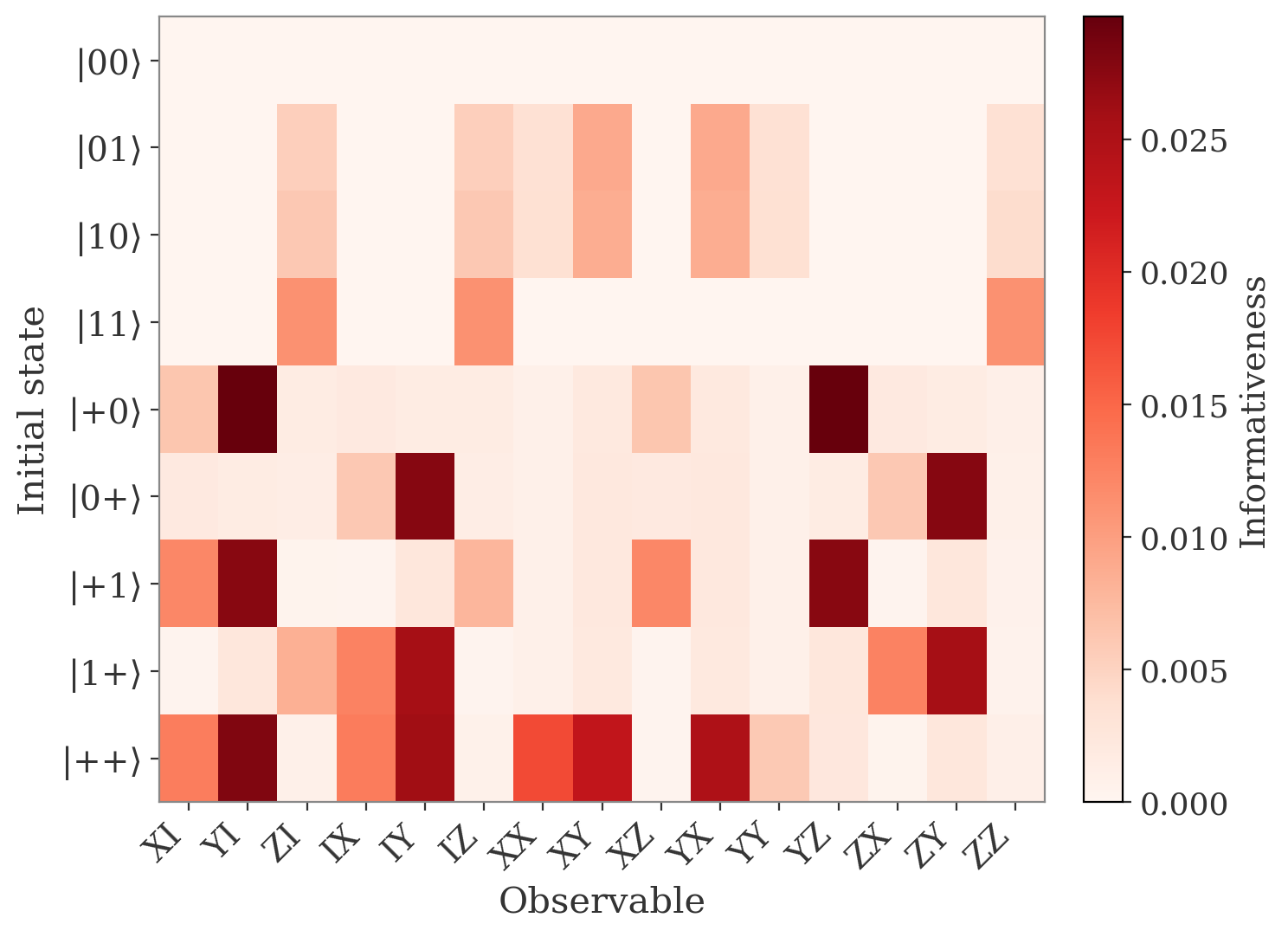}
    \caption{Raw informativeness of every candidate (initial state, observable) pair, computed as the variance of the predicted expectation value across $N=500$ random draws of $(\eta, \phi)$ from the training distribution. Rows: candidate initial states, restricted to tensor products of single-qubit states. Columns: single- and two-qubit Pauli observables (excluding the identity). Darker cells are uninformative; brighter cells are sensitive to the learned parameters.}
    \label{fig:informativeness}
\end{figure}

To keep the per-device measurement budget small we select a compact set of (initial state, observable) pairs that is maximally informative about the device parameters. The candidate set is built from initial states that are tensor products of single-qubit states, which are natural to prepare on superconducting hardware, paired with observables drawn from the single- and two-qubit Pauli operators excluding the identity. For each candidate pair $(\rho_p, O_p)$ we sample $N=500$ random draws of $(\eta, \phi)$ from the training distribution, compute the predicted expectation value for each draw, and assemble these values into an $N$-dimensional signal vector $s_p$. Pairs with high variance across $s_p$ are sensitive to the parameters we wish to learn; pairs whose signal is the same for every $(\eta, \phi)$ provide no information regardless of the device. Figure~\ref{fig:informativeness} reports this raw informativeness as a heatmap over (initial state, observable). Computational-basis initial states are largely uninformative while the most informative cells are superposition states paired with $X$- or $Y$-containing observables.

\begin{figure}[t]
    \centering
    \includegraphics[width=\columnwidth]{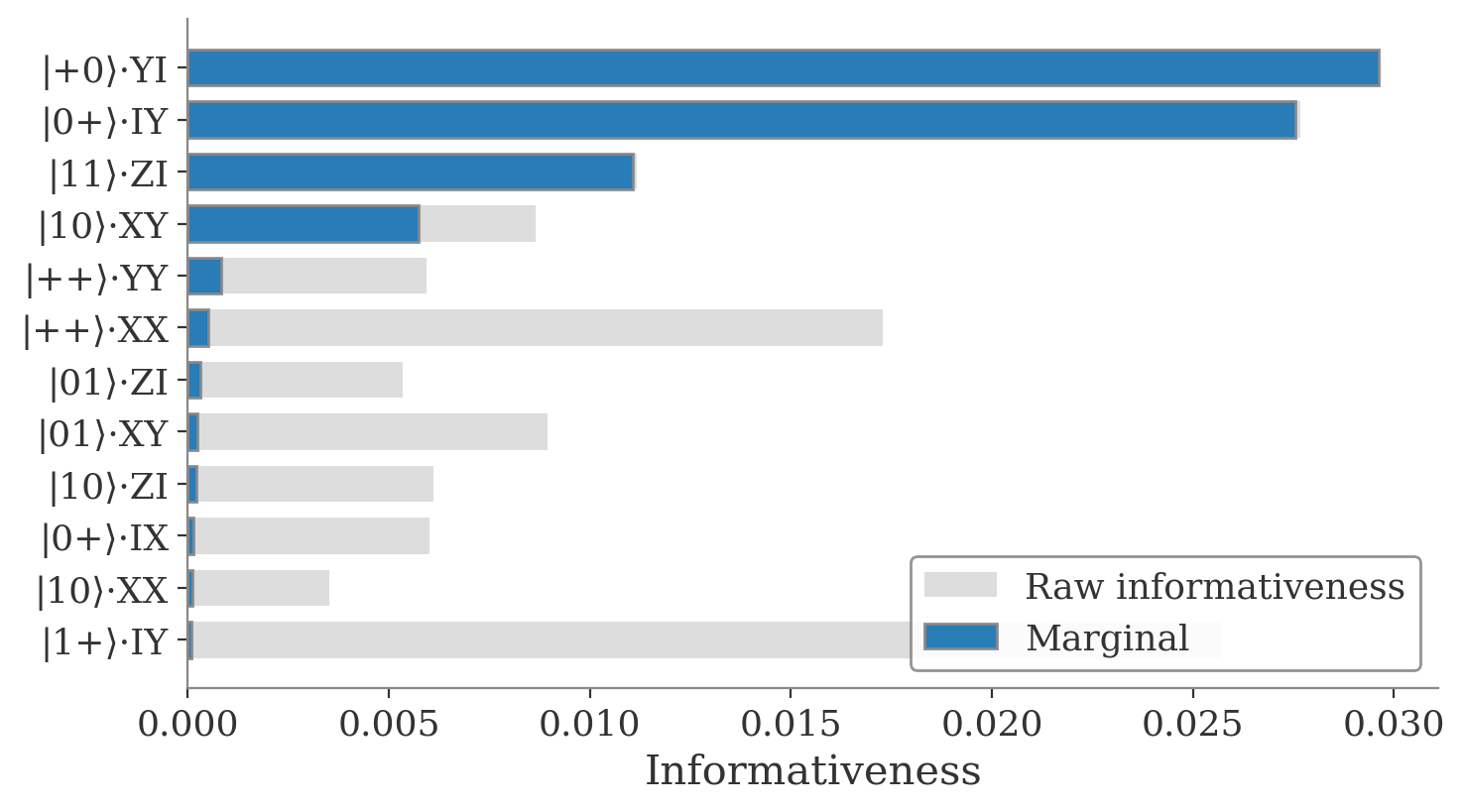}
    \caption{Greedy selection trajectory. Each row corresponds to one (initial state, observable) candidate, ordered by selection (top = first picked); rows below the selected set show the highest-informativeness candidates that were not selected. Gray bars: raw informativeness. Colored bars: marginal informativeness after orthogonalization against previously-selected pairs.}
    \label{fig:marginal}
\end{figure}

We then build the measurement set greedily, using orthogonalization to avoid redundancy. The first selected pair is the one with the largest variance. At each subsequent step, we project every remaining $s_p$ onto the orthogonal complement of the subspace spanned by the previously selected signal vectors and select the pair with the largest residual variance, the marginal informativeness. This guarantees that each new measurement responds to a combination of device parameters not already covered by the existing set. Algorithmically this is QR factorization with column pivoting on the matrix of signal vectors, an approximate greedy maximizer of the submatrix volume $|\det(S^\top S)|$ that has been used in fluid-flow and image-reconstruction settings to build sparse-sensor placements~\cite{manohar2018sparse}. Figure~\ref{fig:marginal} shows the resulting greedy trajectory: each row is a selected pair, with entries in descending order of selection. The gray bar is the pair's raw informativeness and the colored bar is its marginal informativeness after orthogonalization. The first two picks are brand-new information (raw and marginal nearly coincide), while later picks have marginal contributions far below their raw values. Some high-raw candidates near the bottom of the chart are discarded entirely because their marginal contribution is essentially zero. The rapid decay of marginal informativeness motivates terminating selection after a small number of pairs. For the results of Sec.~\ref{sub:coeff_results}--\ref{sub:infid_results} we use $N_M = 7$ measurement pairs.

For adapt-time control pulses we sub-select $N_\phi$ pulses from the random training library of Sec.~\ref{sub:datagen} by farthest-point sampling (FPS)~\cite{eldar1997fps}: after rescaling each flux axis to unit standard deviation across the pool, we seed with the pulse nearest the pool centroid and at each subsequent step add the pulse whose minimum distance to the already-selected set is largest. This spreads the selected pulses evenly across the three-dimensional box of accessible flux configurations. For the results of Sec.~\ref{sub:coeff_results}--\ref{sub:infid_results} we use $N_\phi = 20$, giving a per-device adaptation budget of $N_M \times N_\phi = 140$ expectation values.

\section{Results} \label{sec:results} 
We evaluate HAML on the transmon-coupler-transmon system of Sec.~\ref{sub:system}. The network is trained on 50 simulated devices with 100 randomly sampled control pulses per device, and adapted to 10 held-out devices using the variance-greedy measurement set of Sec.~\ref{sub:measurement} together with 20 adapt-time control pulses selected from the same random training library by farthest-point sampling (Sec.~\ref{sub:measurement}), each measured at a single final evolution time $T$. The SWPT baseline is the second-order approximation evaluated on each held-out device's true parameters. The effective coefficient ground truth is the fidelity-refined target of Sec.~\ref{sub:datagen}.

\subsection{Effective Coefficient Prediction on Held-Out Devices} \label{sub:coeff_results}
\begin{figure}[t]
    \centering
    \includegraphics[width=\columnwidth]{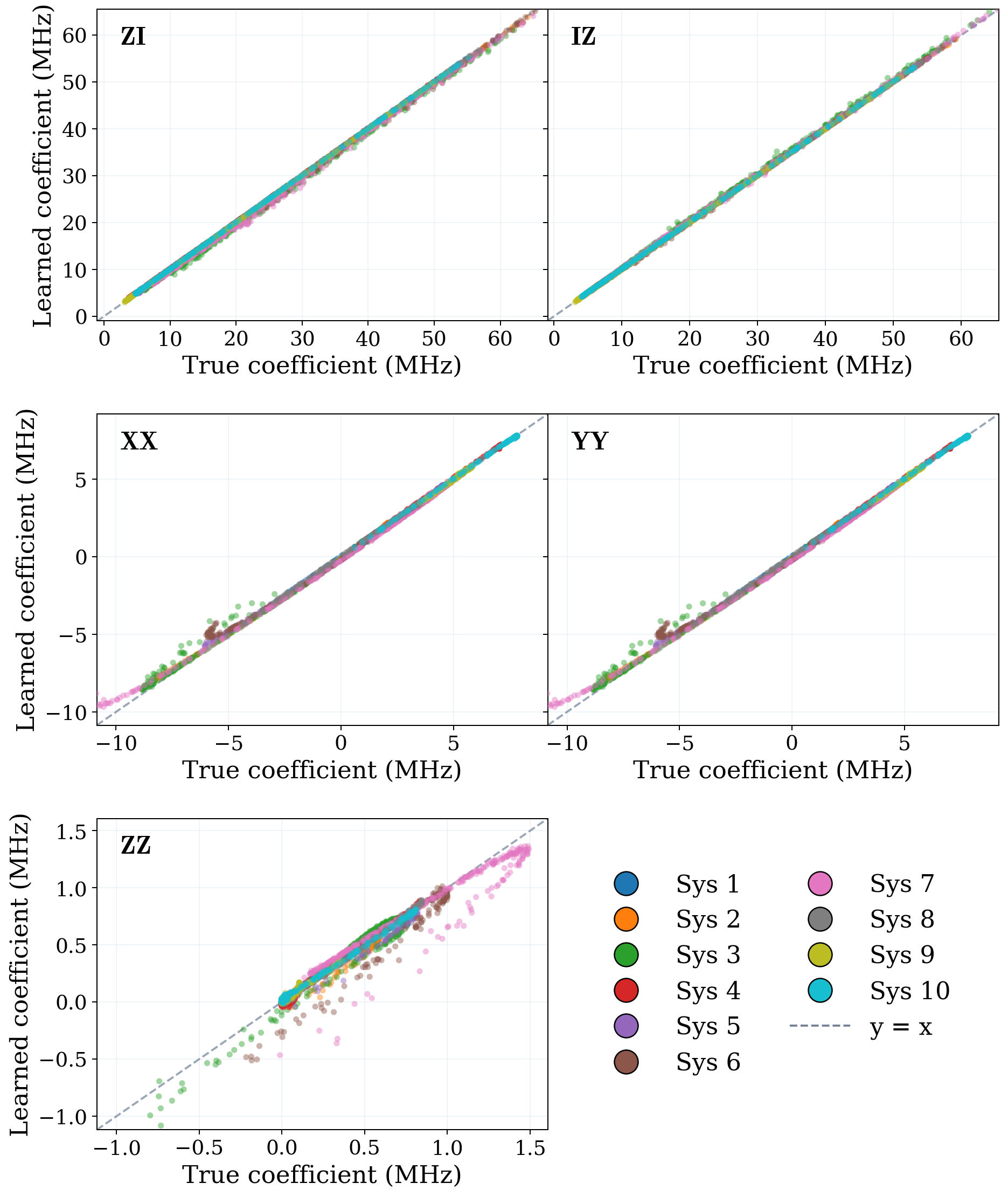}
    \caption{Predicted vs ground-truth projected Pauli coefficients on 10 held-out, adapted devices. Shown are 300 random control points per device. Color encodes device; dashed line is $y=x$.}
    \label{fig:coeff_scatter}
\end{figure}

Figure~\ref{fig:coeff_scatter} shows HAML's predicted effective Pauli coefficients against the effective ground truth for the five terms $\{ZI, IZ, XX, YY, ZZ\}$, evaluated at 300 random control points across each of the 10 held-out devices. The single-qubit detunings $ZI$ and $IZ$ track the $y=x$ diagonal tightly across their full $[0, 60]$~MHz range. The exchange terms $XX, YY$ (range $\pm 10$~MHz) and the parasitic $ZZ$ ($[-1, 1.5]$~MHz) show slightly more spread, particularly in the strong-hybridization corner of each panel, consistent with the held-out set spanning a wide range of qubit-coupler coupling (quantified in Sec.~\ref{sub:hybridization}).

Aggregate accuracy is summarized in Table~\ref{tab:coeff_mae}, alongside the SWPT baseline. Across all five terms, HAML attains a mean absolute error of $0.136$~MHz ($0.58\%$ relative) versus $0.786$~MHz ($4.72\%$ relative) for SWPT---an improvement of roughly $6\times$ in absolute MAE and $8\times$ in relative error.  HAML outperforms SWPT on every term. The gap is largest on the parasitic $ZZ$ term, where HAML's $1.13\%$ relative error is an order of magnitude smaller than SWPT's $11.32\%$; this gap is structural rather than numerical, since second-order Schrieffer-Wolff in the two-level transmon approximation predicts $ZZ = 0$, so the SWPT MAE on $ZZ$ reflects an entire term it omits by construction. On the exchange terms $XX$ and $YY$, HAML reduces the absolute MAE by nearly an order of magnitude ($0.11$~MHz vs $0.92$~MHz); on the single-qubit detunings $ZI$ and $IZ$, HAML is $3$--$5\times$ more accurate.

\begin{table}[t]
    \centering
    \caption{Per-coefficient mean absolute error on the held-out devices.}
    \label{tab:coeff_mae}
    \renewcommand{\arraystretch}{1.15}
    \begin{tabular}{lcccc}
        \hline
        Term & HAML MAE & HAML Rel. & SWPT MAE & SWPT Rel. \\
             & (MHz)    & (\%)      & (MHz)    & (\%)      \\
        \hline
        $ZI$ & 0.259 & 0.39 & 0.872 & 1.32  \\
        $IZ$ & 0.170 & 0.26 & 0.865 & 1.33  \\
        $XX$ & 0.108 & 0.57 & 0.916 & 4.82  \\
        $YY$ & 0.107 & 0.56 & 0.916 & 4.82  \\
        $ZZ$ & 0.036 & 1.13 & 0.362 & 11.32 \\
        \hline
        All & \textbf{0.136} & \textbf{0.58} & \textbf{0.786} & \textbf{4.72} \\
        \hline
    \end{tabular}
    \par\smallskip
    \begin{minipage}{\columnwidth}
        \footnotesize Relative error normalizes each term's MAE by its dynamic range across the held-out set. The \emph{All} row weights all terms and devices equally.
    \end{minipage}
\end{table}
\subsection{The Hybridization Regime Across the Test Ensemble} \label{sub:hybridization}
\begin{figure}[t]
    \centering
    \includegraphics[width=\columnwidth]{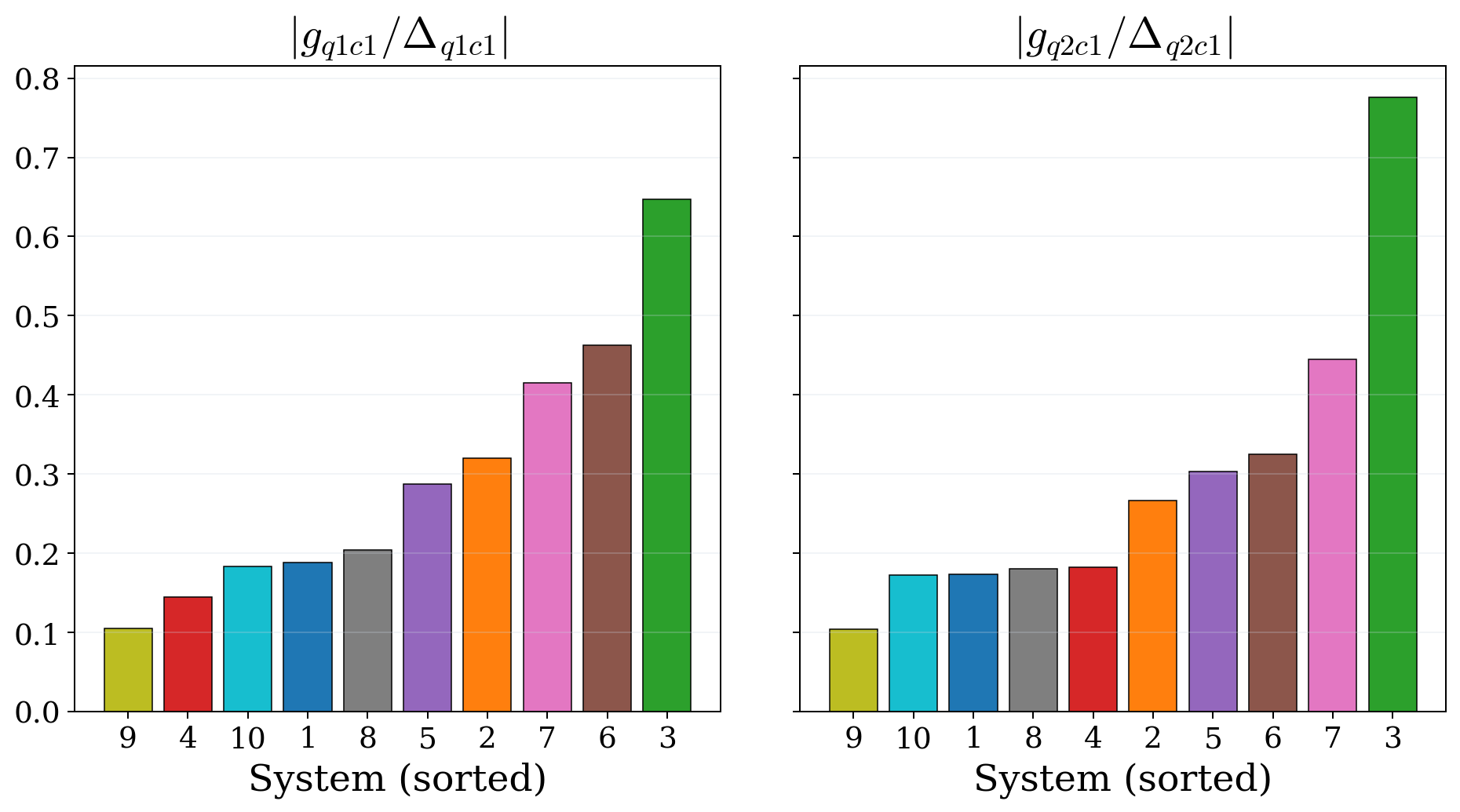}
    \caption{Perturbative expansion ratio $|g_{qi c1}/\Delta_{qi c1}|$ for $i=1$ (left) and $i=2$ (right), across the 10 held-out devices, at the fast-gate operating flux $\Phi_{c1}=1.35$. Bars are sorted ascending; bar color identifies the device.}
    \label{fig:swpt_validity}
\end{figure}

Held-out devices, sampled uniformly from the $\eta$ bounds of Table~\ref{tab:system_params}, sweep through a wide range of qubit-coupler hybridization across their operating range. At low coupler flux ($\Phi_{c1} \approx 0.1$) every device sits in the deep-dispersive regime where SWPT converges well; at the fast-gate operating flux $\Phi_{c1} = 1.35$ they reach varying degrees of strong hybridization, from $|g/\Delta|\!\approx\!0.1$ (Sys 9) to $|g/\Delta|\!\approx\!0.78$ (Sys 3, qubit 2)---comfortably within the regime where the second-order expansion is no longer reliable. Figure~\ref{fig:swpt_validity} reports the perturbative expansion ratio $|g_{qi c1}/\Delta_{qi c1}|$ for both qubits at this operating flux, sorted across the test set.

The same ratio simultaneously controls three things: SWPT's convergence, the leakage of the qubit-like eigenstates into coupler-excited components, and the magnitude of the corrections any 4D effective model cannot recover by construction. To quantify the third point, projecting the full multi-mode propagator $U_{\text{full}}(T)$ onto the qubit-like subspace yields a sub-unitary operator, $U_{\text{proj}}(T) = \Pi U_{\text{full}}(T) \Pi$, whose deviation from unitarity is exactly the leakage out of that subspace during evolution. Comparing any 4D effective unitary $U_{\text{eff}} = \exp(-i H_{\text{eff}} T)$ against $U_{\text{proj}}(T)$ thus carries an irreducible residual that we refer to as the \emph{infidelity floor}: it depends on the projection itself, not on the model. We denote the infidelity floor by $I_\text{floor}\equiv 1-F_\text{floor}$, where $F_\text{floor}$ is the standard process fidelity between $U_{\text{eff}}$ and $U_{\text{proj}}(T)$. HAML and SWPT both inherit this floor at strong hybridization; SWPT additionally diverges from it as its expansion degrades, while HAML, trained against the exact dressed-state coefficients, stays near it. Sec.~\ref{sub:infid_results} makes the asymmetric divergence quantitative.

\subsection{HAML vs SWPT Across the Operating Range} \label{sub:infid_results}

\begin{figure}[t]
    \centering
    \includegraphics[width=\columnwidth]{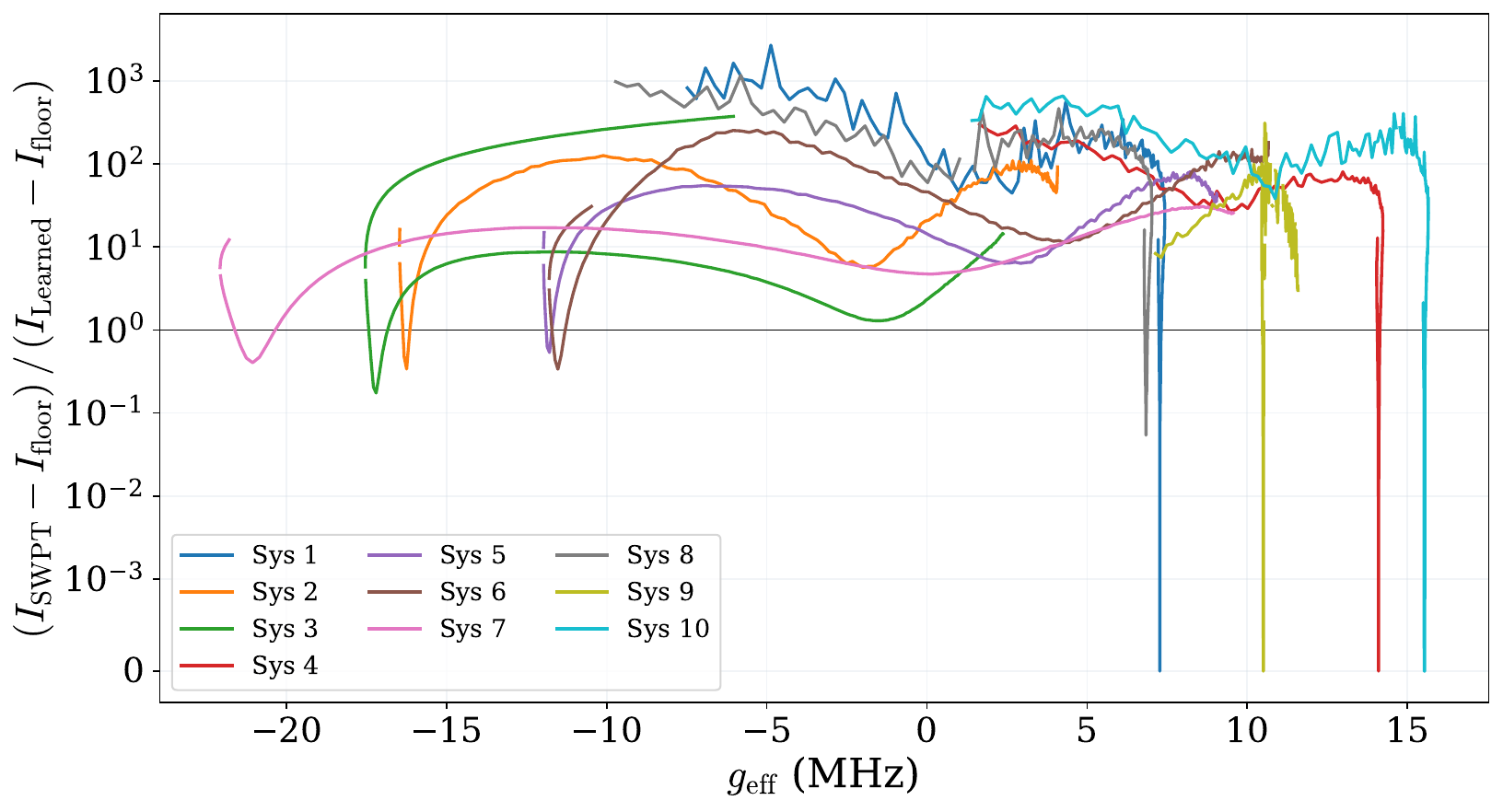}
     \caption{Infidelity gain of HAML over SWPT, $(I_{\text{SWPT}} - I_\text{Floor})/(I_{\text{Learned}} - I_\text{floor})$, with fidelity floor $I_\text{floor}$, as a function of effective coupling $g_{\text{eff}}$, per held-out device.}
    \label{fig:infid_comparison}
\end{figure}

For each held-out device we sweep the operating flux and compare the HAML and SWPT projected-unitary fidelities against the true $U_{\text{proj}}(T)$. Figure~\ref{fig:infid_comparison} plots the ratio of excess infidelities $(I_{\text{SWPT}} - I_{\text{floor}})/(I_{\text{HAML}} - I_{\text{floor}})$ as a function of effective coupling $g_{\text{eff}}$, on a logarithmic scale, where $I_{\text{floor}}$ is the irreducible infidelity floor defined in Sec.~\ref{sub:hybridization} and values above unity indicate HAML wins. After subtracting this floor, HAML reduces the model infidelity by $10\times$--$1000\times$ relative to SWPT in most systems. Aggregating across the held-out set, HAML's mean excess infidelity is $1.1\times 10^{-5}$, compared with $4.3\times 10^{-4}$ for SWPT---a factor of roughly $40$. Increasing $\Phi_{c_1}$ drives $g_{\text{eff}}$ from small positive values at low flux (where SWPT performs best) toward strongly negative values at high flux (where SWPT's perturbative expansion is least accurate). The narrow operating points where SWPT marginally outperforms HAML are not regions of genuine SWPT accuracy: they correspond to fluxes where SWPT's drifting coefficient curves momentarily cross through the ground truth on their way past it, as visible in the term-by-term sweep in Fig.~\ref{fig:coeff_sweep} (Appendix~\ref{appendix:coeff_sweep}).

SWPT's accuracy is bounded analytically by the order parameter $\epsilon = g_{qc}/\Delta_{qc}$ and degrades once the neglected $\mathcal{O}(\epsilon^3)$ terms stop being small, and extending to higher order requires deriving a nontrivial generator for the block-diagonalization. HAML is not tied to any such expansion. Both stages operate on the full multi-mode dynamics projected onto the qubit subspace: training fits effective coefficients to the projected unitary (Sec.~\ref{sub:datagen}), and adaptation fits device parameters to qubit-subspace observables (Sec.~\ref{sub:adaptation}). HAML therefore captures higher-order corrections across all terms---including the parasitic $ZZ$ interaction that second-order SWPT omits entirely---and can be interpreted as a data-driven analog of a higher-order reduction, bypassing the nontrivial analytic derivation it would otherwise require.

\section{Conclusion} \label{sec:conclusion}
From a simulated device ensemble, HAML learns the reduction from full multi-mode quantum dynamics to an effective qubit Hamiltonian, and adapts to a new device from a handful of measurements. On a transmon-coupler-transmon benchmark, the trained model reduces effective-coefficient MAE by $\sim$$6\times$ and projected-unitary excess infidelity by a factor of $\sim$$40\times$ over second-order SWPT, with the gap concentrated in the near-resonance regime relevant to fast two-qubit gates. Once trained, adaptation to a new device takes seconds on a single CPU. HAML also carries practical advantages for device calibration. Because adaptation operates entirely on qubit-subspace observables, HAML can be deployed on architectures where the coupler lacks dedicated readout without first running the indirect spectroscopy needed to back out bare mode parameters for a perturbative reduction.

Several extensions follow from our current iteration of HAML. First, retraining with three levels per mode rather than two would capture leakage into the second excited state and an additional contribution to the parasitic $ZZ$ interaction strength. Second, extending to time-varying pulses would allow the same pipeline to predict effective coefficients along a trajectory, enabling characterization of time-dependent device characteristics such as pulse distortion~\cite{genois2021}. Third, scaling to multi-coupler architectures is a natural next step, as SWPT becomes more analytically complex in this setting. Finding the perturbative ansatz becomes difficult, while HAML's cost scales only with the number of device parameters $\eta$ and remains analytically straightforward. Finally, characterizing performance under realistic measurement noise is necessary before hardware deployment, where measurement set optimizations that balance variance maximization against noise insensitivity will be especially valuable.

Broadly, our results suggest that effective Hamiltonian reduction---traditionally a pen-and-paper exercise in symbolic manipulation---can instead be solved by training a network on simulated examples whenever a high-fidelity digital twin is available. As superconducting architectures grow in mode count and operating bandwidth, and as detailed device simulators become more common~\cite{Bayraktar2023}, data-driven reductions are poised to become increasingly attractive relative to the case-by-case derivations they replace.

\section*{Acknowledgments} \label{sec:ackn}
FTC is the Chief Scientist for Quantum Software at Infleqtion.

This work is funded in part by the STAQ project under award NSF Phy-232580; in part by the US Department of Energy Office of Advanced Scientific Computing Research, Accelerated Research for Quantum Computing Program (Award Number DE-SC0025509); in part by the NSF Quantum Leap Challenge Institute for Hybrid Quantum Architectures and Networks (NSF Award 2016136); in part by the NSF National Virtual Quantum Laboratory program; in part based upon work supported by the U.S. Department of Energy, Office of Science, National Quantum Information Science Research Centers; in part by the Army Research Office under Grant Number W911NF-23-1-0077. The views and conclusions contained in this document are those of the authors and should not be interpreted as representing the official policies, either expressed or implied, of the U.S. Government. The U.S. Government is authorized to reproduce and distribute reprints for Government purposes notwithstanding any copyright notation herein.

\appendices
\numberwithin{equation}{section}
\numberwithin{figure}{section}
\numberwithin{table}{section}

\section{Convergence} \label{appendix:convergence}
\begin{figure}[h]
    \centering
    \includegraphics[width=\columnwidth]{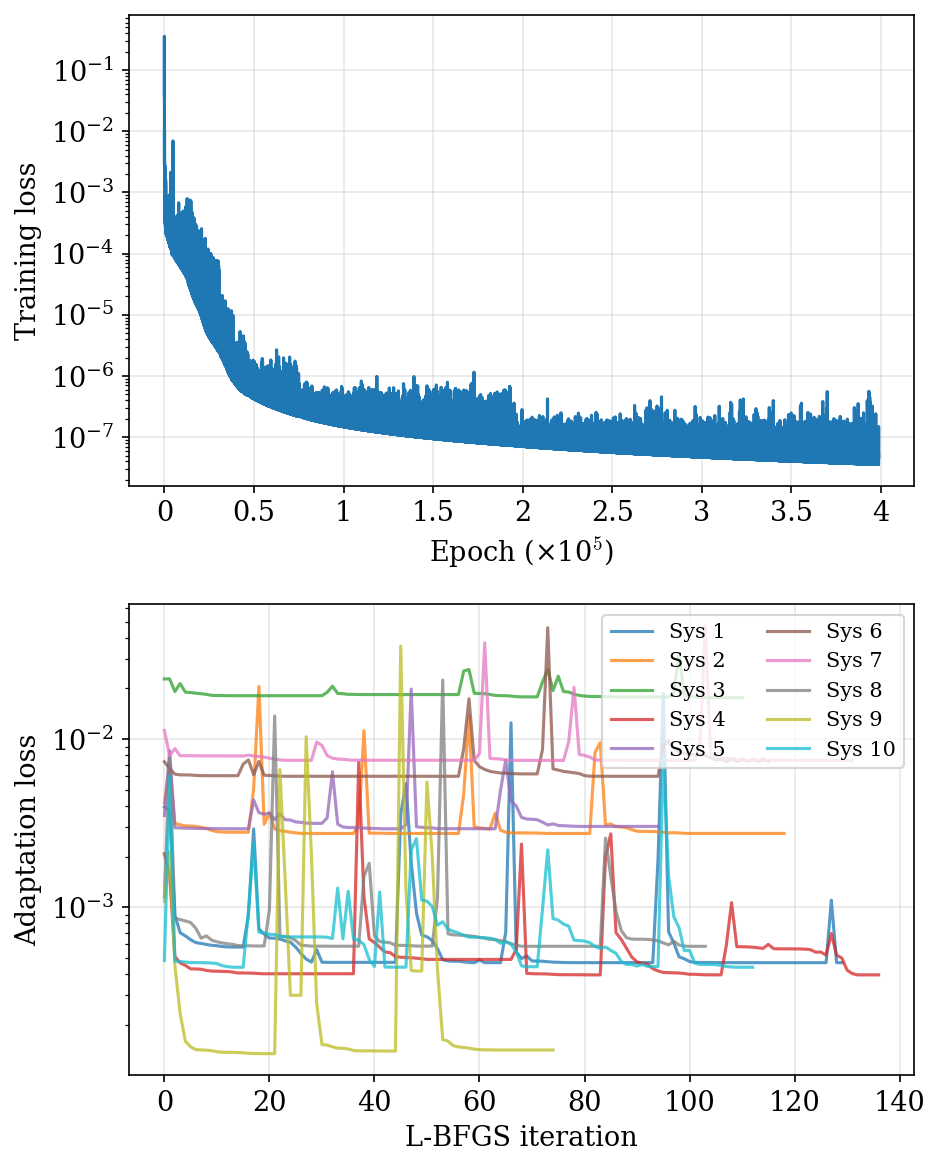}
    \caption{Training loss vs.\ epoch (left) and L-BFGS adaptation loss vs.\ iteration for each held-out device (right), both on a log scale.}
    \label{fig:loss}
\end{figure}

Figure~\ref{fig:loss} shows the training loss across epochs and the L-BFGS adaptation loss for each of the 10 held-out devices. Training drives the projected-coefficient MSE down by roughly seven orders of magnitude, with the learning-rate scheduler taking over once the curve plateaus. Adaptation converges in fewer than 140 total L-BFGS iterations per device, summed over 5 random restarts; the visible spikes correspond to restart events at which $\eta_{\text{pred}}$ is reseeded uniformly within the bounds of Table~\ref{tab:system_params}.

\FloatBarrier
\section{System Parameters} \label{appendix:system_params}

The two transmons are held fixed across the device ensemble; only the coupler-related parameters $\eta$ vary. Table~\ref{tab:system_params} lists the fixed qubit values, the sampling ranges for $\eta$ used to generate the training and held-out devices, and the control flux ranges spanned by the pulse library.

\begin{table}[h]
    \centering
    \caption{System parameters.}
    \label{tab:system_params}
    \renewcommand{\arraystretch}{1.35}
    \begin{tabular}{lll}
        \hline
        \multicolumn{3}{l}{\textit{Fixed qubit parameters}} \\
        \hline
        $E_J^{0}$ (q1, q2)              & 20.0 & GHz \\
        $E_C$ (q1, q2)                  & 0.25 & GHz \\
        \hline
        \multicolumn{3}{l}{\textit{Variable device parameters $\eta$ (uniform sampling)}} \\
        \hline
        $E_J^{0,c1}$                    & $[23.0,\, 28.0]$    & GHz \\
        $E_C^{c1}$                      & $[0.28,\, 0.32]$    & GHz \\
        $E_C^{q1,c1}$                   & $[0.015,\, 0.025]$  & GHz \\
        $E_C^{q2,c1}$                   & $[0.015,\, 0.025]$  & GHz \\
        $E_C^{q1,q2}$                   & $[0.002,\, 0.004]$  & GHz \\
        \hline
        \multicolumn{3}{l}{\textit{Control flux ranges (constant-pulse library)}} \\
        \hline
        $\phi_{ext}^{q1}$               & $[0.0,\, 0.5]$      & rad \\
        $\phi_{ext}^{q2}$               & $[0.0,\, 0.5]$      & rad \\
        $\phi_{ext}^{c1}$               & $[0.1,\, 1.35]$     & rad \\
        \hline
    \end{tabular}
\end{table}

\FloatBarrier
\section{Training and Adaptation Hyperparameters} \label{appendix:hyperparams}

Table~\ref{tab:hyperparams} reports the network architecture, optimizer settings, dataset sizes, and L-BFGS adaptation parameters used to produce the results of Sec.~\ref{sub:coeff_results}--\ref{sub:infid_results}.

\begin{table}[h]
    \centering
    \caption{Network, training, and adaptation hyperparameters.}
    \label{tab:hyperparams}
    \begin{tabular}{ll}
        \hline
        \multicolumn{2}{l}{\textit{Network architecture}} \\
        \hline
        Hidden layers                   & 3 \\
        Hidden dimension                & 64 \\
        Activation                      & SiLU \\
        \hline
        \multicolumn{2}{l}{\textit{Training}} \\
        \hline
        Optimizer                       & Adam \\
        Initial learning rate           & $2 \times 10^{-2}$ \\
        Scheduler patience (epochs)     & 500 \\
        LR decay factor                 & 0.5 \\
        Minimum learning rate           & $1 \times 10^{-7}$ \\
        Early-stopping patience (epochs)& 600 \\
        Batch size                      & full ($50 \times 100$) \\
        Best epoch (this run)           & 398{,}626 \\
        \hline
        \multicolumn{2}{l}{\textit{Dataset}} \\
        \hline
        Training devices                & 50 \\
        Held-out devices                & 10 \\
        Training control pulses per device & 100 \\
        Adapt control pulses per device & 20 \\
        Final evolution time $T$        & 1~ns \\
        Time snapshots (training only)  & 10 \\
        \hline
        \multicolumn{2}{l}{\textit{Adaptation}} \\
        \hline
        Optimizer                       & L-BFGS-B \\
        Bounds                          & $\eta$ ranges (Table~\ref{tab:system_params}) \\
        Max iterations per restart      & 100 \\
        Random restarts                 & 5 \\
        Initialization                  & uniform within $\eta$ bounds \\
        Evaluation time                 & $T$ only (no intermediate times) \\
        \hline
    \end{tabular}
\end{table}

\FloatBarrier
\section{Per-coefficient flux sweep} \label{appendix:coeff_sweep}

\begin{figure}[th]
    \centering
    \includegraphics[width=\columnwidth]{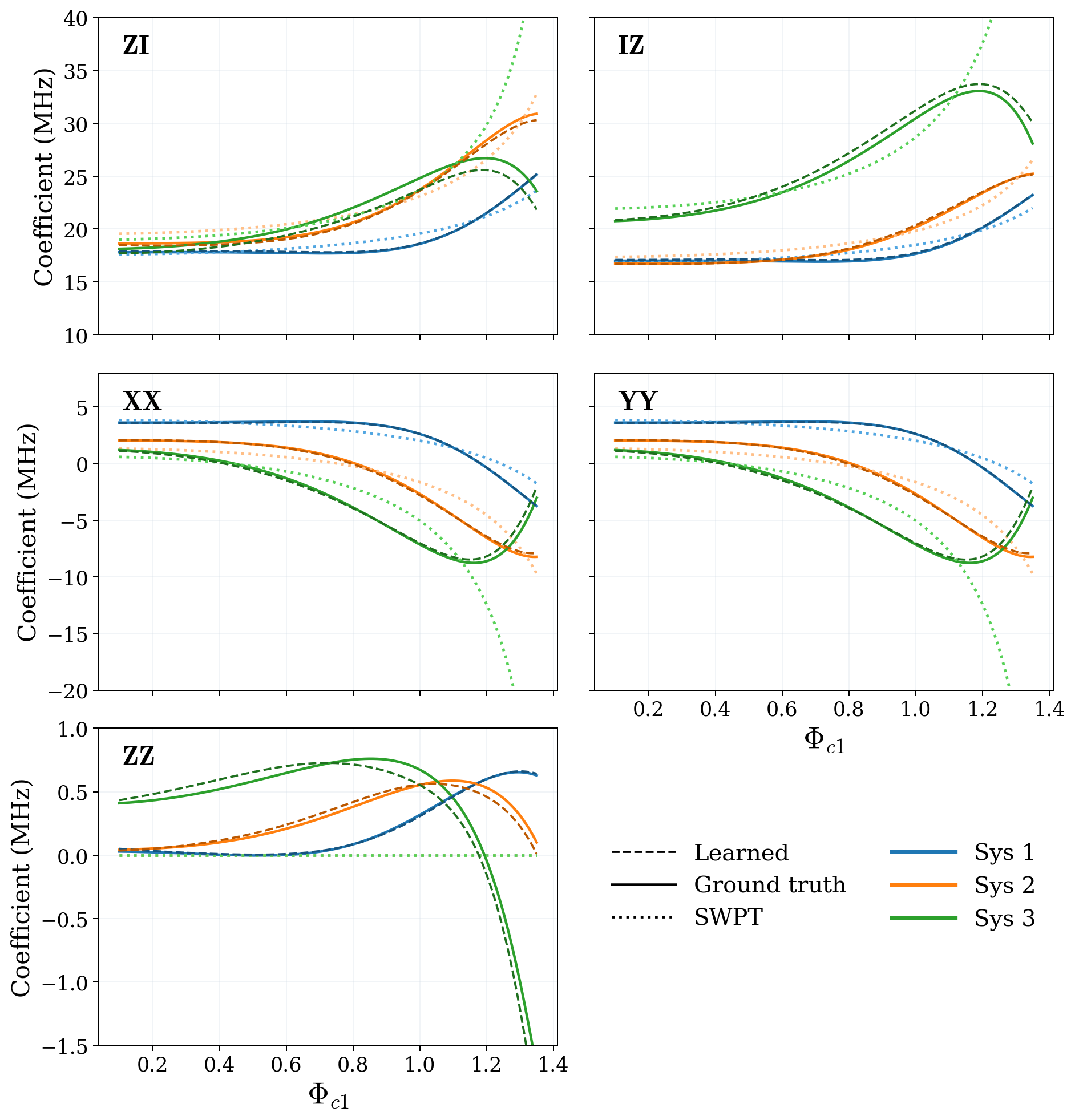}
    \caption{Projected-Pauli coefficients vs.\ coupler flux $\Phi_{c_1}$ for three held-out devices. Dashed: HAML prediction; solid: fidelity-refined ground truth; dotted: SWPT. Qubit fluxes held at $\phi_{q_1}=\phi_{q_2}=0.25$.}
    \label{fig:coeff_sweep}
\end{figure}

Figure~\ref{fig:coeff_sweep} compares HAML, SWPT, and the fidelity-refined ground truth across the full coupler-flux operating range for three representative held-out devices, with one panel per projected-Pauli term. HAML's prediction (dashed) tracks the ground-truth curve (solid) closely across the full range for all five terms; small residual gaps appear for the most hybridized device (Sys 3) near the high-flux endpoint, most visibly on $ZZ$. SWPT (dotted) tracks the truth in the low-flux dispersive regime but drifts progressively further as the coupler is tuned through resonance---visible on $ZI$/$IZ$ and on $XX$/$YY$ at large $\Phi_{c_1}$, and on $ZZ$ across the entire sweep, since its second-order transmon truncation cannot reproduce the parasitic $ZZ$ term at all.

\FloatBarrier
\section{Runtime} \label{appendix:runtime}

Wall-clock costs for the run reported in Secs.~\ref{sub:coeff_results}--\ref{sub:infid_results}, measured on a 2-core CPU node of the NERSC Perlmutter system (no GPU):
\begin{itemize}
    \item \textbf{Training-data generation:} $0.565$~s per device ($100$ pulses), or $\sim\!28$~s for the full 50-device set, single-threaded.
    \item \textbf{Training:} $972$~min (one-time), to the best epoch reported in Table~\ref{tab:hyperparams}.
    \item \textbf{Adaptation (per held-out device):} $6.09$~s for the full 5-restart L-BFGS loop, or $\sim\!53$~ms per iteration.
\end{itemize}
The per-device adaptation cost is the operationally relevant number: once trained, HAML specializes to a new device in seconds without requiring a GPU.

\clearpage
\bibliographystyle{IEEEtran}
\bibliography{bibs/main,bibs/arxiv}

@article{petterssonfors2024,
  title={Comprehensive explanation of {ZZ} coupling in superconducting qubits},
  author={Fors, Simon Pettersson and Fern{\'a}ndez-Pend{\'a}s, Jorge and Kockum, Anton Frisk},
  journal={arXiv preprint arXiv:2408.15402},
  year={2024}
}

@misc{schorling2025metalearning,
      title={Meta-learning characteristics and dynamics of quantum systems}, 
      author={Lucas Schorling and Pranav Vaidhyanathan and Jonas Schuff and Miguel J. Carballido and Dominik Zumbühl and Gerard Milburn and Florian Marquardt and Jakob Foerster and Michael A. Osborne and Natalia Ares},
      year={2025},
      eprint={2503.10492},
      archivePrefix={arXiv},
      primaryClass={quant-ph},
      url={https://arxiv.org/abs/2503.10492}, 
}

@article{wu2026,
    author = "Wu, Xuntao and Yan, Haoxiong and Andersson, Gustav and Anferov, Alexander and Conner, Christopher R. and Joshi, Yash J. and Karimi, Bayan and King, Amber M. and Li, Shiheng and Malc, Howard L. and Miller, Jacob M. and Mishra, Harsh and Qiao, Hong and Ryu, Minseok and Shi, Jian and Cleland, Andrew N.",
    title = "{Efficient $n$-qubit entangling operations via a superconducting quantum router}",
    eprint = "2604.15432",
    archivePrefix = "arXiv",
    primaryClass = "quant-ph",
    month = "4",
    year = "2026"
}

@misc{UnitaryTransformations.jl,
  author = {Karle, Volker},
  title = {{UnitaryTransformations.jl: Symbolic Unitary Transformations for Quantum Hamiltonians}},
  url = {https://github.com/volkerkarle/UnitaryTransformations.jl},
  year = {2025}
}

@article{yan2018,
  title = {Tunable Coupling Scheme for Implementing High-Fidelity Two-Qubit Gates},
  author = {Yan, Fei and Krantz, Philip and Sung, Youngkyu and Kjaergaard, Morten and Campbell, Daniel L. and Orlando, Terry P. and Gustavsson, Simon and Oliver, William D.},
  journal = {Phys. Rev. Appl.},
  volume = {10},
  issue = {5},
  pages = {054062},
  numpages = {9},
  year = {2018},
  month = {Nov},
  publisher = {American Physical Society},
  doi = {10.1103/PhysRevApplied.10.054062},
  url = {https://link.aps.org/doi/10.1103/PhysRevApplied.10.054062}
}

@article{blais2021circuit,
  title = {Circuit quantum electrodynamics},
  volume = {93},
  ISSN = {1539-0756},
  url = {http://dx.doi.org/10.1103/RevModPhys.93.025005},
  DOI = {10.1103/revmodphys.93.025005},
  number = {2},
  journal = {Reviews of Modern Physics},
  publisher = {American Physical Society (APS)},
  author = {Blais,  Alexandre and Grimsmo,  Arne L. and Girvin,  S. M. and Wallraff,  Andreas},
  year = {2021},
  month = may 
}

@inproceedings{chen2018neural,
    author = {Chen, Ricky T. Q. and Rubanova, Yulia and Bettencourt, Jesse and Duvenaud, David},
    title = {Neural ordinary differential equations},
    year = {2018},
    publisher = {Curran Associates Inc.},
    address = {Red Hook, NY, USA},
    booktitle = {Proceedings of the 32nd International Conference on Neural Information Processing Systems},
    pages = {6572–6583},
    numpages = {12},
    location = {Montr\'{e}al, Canada},
    series = {NIPS'18},
    doi={10.5555/3327757.3327764},
    url={https://dl.acm.org/doi/10.5555/3327757.3327764}
}

@article{vandamme2024advanced,
  title = {Advanced {CMOS} manufacturing of superconducting qubits on 300 mm wafers},
  volume = {634},
  ISSN = {1476-4687},
  url = {http://dx.doi.org/10.1038/s41586-024-07941-9},
  DOI = {10.1038/s41586-024-07941-9},
  number = {8032},
  journal = {Nature},
  publisher = {Springer Science and Business Media LLC},
  author = {Van Damme,  J. and Massar,  S. and Acharya,  R. and Ivanov,  Ts. and Perez Lozano,  D. and Canvel,  Y. and Demarets,  M. and Vangoidsenhoven,  D. and Hermans,  Y. and Lai,  J. G. and Vadiraj,  A. M. and Mongillo,  M. and Wan,  D. and De Boeck,  J. and Potočnik,  A. and De Greve,  K.},
  year = {2024},
  month = {sep},
  pages = {74–79}
}

@article{burnett2019decoherence,
  title = {Decoherence benchmarking of superconducting qubits},
  volume = {5},
  ISSN = {2056-6387},
  url = {http://dx.doi.org/10.1038/s41534-019-0168-5},
  DOI = {10.1038/s41534-019-0168-5},
  number = {1},
  journal = {npj Quantum Information},
  publisher = {Springer Science and Business Media LLC},
  author = {Burnett,  Jonathan J. and Bengtsson,  Andreas and Scigliuzzo,  Marco and Niepce,  David and Kudra,  Marina and Delsing,  Per and Bylander,  Jonas},
  year = {2019},
  month = {jun} 
}

@article{goldschmidt2021bilinear,
  title={Bilinear dynamic mode decomposition for quantum control},
  author={Goldschmidt, Andy and Kaiser, Eurika and Dubois, Jonathan L and Brunton, Steven L and Kutz, J Nathan},
  journal={New Journal of Physics},
  volume={23},
  number={3},
  pages={033035},
  year={2021},
  publisher={IOP Publishing},
  doi={10.1088/1367-2630/abe972},
  url={https://iopscience.iop.org/article/10.1088/1367-2630/abe972/meta}
}

@article{gambetta2016,
  title = {Universal Gate for Fixed-Frequency Qubits via a Tunable Bus},
  author = {McKay, David C. and Filipp, Stefan and Mezzacapo, Antonio and Magesan, Easwar and Chow, Jerry M. and Gambetta, Jay M.},
  journal = {Phys. Rev. Appl.},
  volume = {6},
  issue = {6},
  pages = {064007},
  numpages = {10},
  year = {2016},
  month = {Dec},
  publisher = {American Physical Society},
  doi = {10.1103/PhysRevApplied.6.064007},
  url = {https://link.aps.org/doi/10.1103/PhysRevApplied.6.064007}
}

@article{wu2024,
  title = {Modular Quantum Processor with an All-to-All Reconfigurable Router},
  author = {Wu, Xuntao and Yan, Haoxiong and Andersson, Gustav and Anferov, Alexander and Chou, Ming-Han and Conner, Christopher R. and Grebel, Joel and Joshi, Yash J. and Li, Shiheng and Miller, Jacob M. and Povey, Rhys G. and Qiao, Hong and Cleland, Andrew N.},
  journal = {Phys. Rev. X},
  volume = {14},
  issue = {4},
  pages = {041030},
  numpages = {22},
  year = {2024},
  month = {Nov},
  publisher = {American Physical Society},
  doi = {10.1103/PhysRevX.14.041030},
  url = {https://link.aps.org/doi/10.1103/PhysRevX.14.041030}
}

@article{zhang2023characterization,
  title={Characterization of tunable coupler without a dedicated readout resonator in superconducting circuits},
  author={Zhang, Chi and Wang, Tian-Le and Guo, Liang-Liang and Yang, Xiao-Yan and Yang, Xin-Xin and Duan, Peng and Jia, Zhi-Long and Kong, Wei-Cheng and Guo, Guo-Ping},
  journal={Applied Physics Letters},
  volume={122},
  number={2},
  year={2023},
  publisher={AIP Publishing},
  doi={https://doi.org/10.1063/5.0135219},
  url={https://pubs.aip.org/aip/apl/article/122/2/024001/2876150/Characterization-of-tunable-coupler-without-a}
}

@article{arute2019quantum,
  title={Quantum supremacy using a programmable superconducting processor},
  author={Arute, Frank and Arya, Kunal and Babbush, Ryan and Bacon, Dave and Bardin, Joseph C and Barends, Rami and Biswas, Rupak and Boixo, Sergio and Brandao, Fernando GSL and Buell, David A and others},
  journal={nature},
  volume={574},
  number={7779},
  pages={505--510},
  year={2019},
  publisher={Nature Publishing Group UK London},
  doi={https://doi.org/10.1038/s41586-019-1666-5},
  url={https://doi.org/10.1038/s41586-019-1666-5}
}

@article{google2025quantum,
  title={Quantum error correction below the surface code threshold},
  journal={Nature},
  volume={638},
  number={8052},
  pages={920--926},
  year={2025},
  publisher={Nature Publishing Group UK London},
  doi={https://doi.org/10.1038/s41586-024-08449-y},
  url={https://doi.org/10.1038/s41586-024-08449-y}
}

@article{stehlik2021,
  title = {Tunable Coupling Architecture for Fixed-Frequency Transmon Superconducting Qubits},
  author = {Stehlik, J. and Zajac, D. M. and Underwood, D. L. and Phung, T. and Blair, J. and Carnevale, S. and Klaus, D. and Keefe, G. A. and Carniol, A. and Kumph, M. and Steffen, Matthias and Dial, O. E.},
  journal = {Phys. Rev. Lett.},
  volume = {127},
  issue = {8},
  pages = {080505},
  numpages = {6},
  year = {2021},
  month = {Aug},
  publisher = {American Physical Society},
  doi = {10.1103/PhysRevLett.127.080505},
  url = {https://link.aps.org/doi/10.1103/PhysRevLett.127.080505}
}

@article{li2020,
  title = {Tunable Coupler for Realizing a Controlled-Phase Gate with Dynamically Decoupled Regime in a Superconducting Circuit},
  author = {Li, X. and Cai, T. and Yan, H. and Wang, Z. and Pan, X. and Ma, Y. and Cai, W. and Han, J. and Hua, Z. and Han, X. and Wu, Y. and Zhang, H. and Wang, H. and Song, Yipu and Duan, Luming and Sun, Luyan},
  journal = {Phys. Rev. Appl.},
  volume = {14},
  issue = {2},
  pages = {024070},
  numpages = {14},
  year = {2020},
  month = {Aug},
  publisher = {American Physical Society},
  doi = {10.1103/PhysRevApplied.14.024070},
  url = {https://link.aps.org/doi/10.1103/PhysRevApplied.14.024070}
}

@article{wu2025,
  title = {Mitigating cosmic-ray-like correlated events with a modular quantum processor},
  author = {Wu, Xuntao and Joshi, Yash J. and Yan, Haoxiong and Andersson, Gustav and Anferov, Alexander and Conner, Christopher R. and Karimi, Bayan and King, Amber M. and Li, Shiheng and Malc, Howard L. and Miller, Jacob M. and Mishra, Harsh and Qiao, Hong and Ryu, Minseok and Xing, Siyuan and Shi, Jian and Cleland, Andrew N.},
  journal = {Phys. Rev. Appl.},
  volume = {24},
  issue = {4},
  pages = {044022},
  numpages = {12},
  year = {2025},
  month = {Oct},
  publisher = {American Physical Society},
  doi = {10.1103/4ctq-r6w6},
  url = {https://link.aps.org/doi/10.1103/4ctq-r6w6}
}

@article{li2025,
  title = {High-precision pulse calibration of tunable couplers for high-fidelity two-qubit gates in superconducting quantum processors},
  author = {Li, Tian-Ming and Zhang, Jia-Chi and Chen, Bing-Jie and Huang, Kaixuan and Liu, Hao-Tian and Xiao, Yong-Xi and Deng, Cheng-Lin and Liang, Gui-Han and Chen, Chi-Tong and Liu, Yu and Li, Hao and Bao, Zhen-Ting and Zhao, Kui and Xu, Yueshan and Li, Li and He, Yang and Liu, Zheng-He and Yu, Yi-Han and Zhou, Si-Yun and Liu, Yan-Jun and Song, Xiaohui and Zheng, Dongning and Xiang, Zhongcheng and Shi, Yun-Hao and Xu, Kai and Fan, Heng},
  journal = {Phys. Rev. Appl.},
  volume = {23},
  issue = {2},
  pages = {024059},
  numpages = {16},
  year = {2025},
  month = {Feb},
  publisher = {American Physical Society},
  doi = {10.1103/PhysRevApplied.23.024059},
  url = {https://link.aps.org/doi/10.1103/PhysRevApplied.23.024059}
}

@article{zhang2025,
  title = {Characterization and optimization of tunable couplers via adiabatic control in superconducting circuits},
  author = {Zhang, Xuan and Zhang, Xu and Chen, Changling and Tang, Kai and Yi, Kangyuan and Luo, Kai and Xie, Zheshu and Chen, Yuanzhen and Yan, Tongxing},
  journal = {Phys. Rev. Appl.},
  volume = {24},
  issue = {3},
  pages = {034003},
  numpages = {7},
  year = {2025},
  month = {Sep},
  publisher = {American Physical Society},
  doi = {10.1103/wgqf-8rcy},
  url = {https://link.aps.org/doi/10.1103/wgqf-8rcy}
}

@article{bravyi2011swpt,
    title = {Schrieffer–Wolff transformation for quantum many-body systems},
    journal = {Annals of Physics},
    volume = {326},
    number = {10},
    pages = {2793-2826},
    year = {2011},
    issn = {0003-4916},
    doi = {https://doi.org/10.1016/j.aop.2011.06.004},
    url = {https://www.sciencedirect.com/science/article/pii/S0003491611001059},
    author = {Sergey Bravyi and David P. DiVincenzo and Daniel Loss}
}

@article{li2023imode,
  title = {Metalearning Generalizable Dynamics from Trajectories},
  author = {Li, Qiaofeng and Wang, Tianyi and Roychowdhury, Vwani and Jawed, M. K.},
  journal = {Phys. Rev. Lett.},
  volume = {131},
  issue = {6},
  pages = {067301},
  numpages = {6},
  year = {2023},
  month = {Aug},
  publisher = {American Physical Society},
  doi = {10.1103/PhysRevLett.131.067301},
  url = {https://link.aps.org/doi/10.1103/PhysRevLett.131.067301}
}

@inproceedings{finn2017maml,
  author = {Finn, Chelsea and Abbeel, Pieter and Levine, Sergey}, title = {Model-agnostic meta-learning for fast adaptation of deep networks}, year = {2017}, publisher = {JMLR.org}, booktitle = {Proceedings of the 34th International Conference on Machine Learning - Volume 70}, pages = {1126–1135}, numpages = {10}, location = {Sydney, NSW, Australia}, series = {ICML'17} }

@inproceedings{zintgraf2019cavia,
  title = {Fast Context Adaptation via Meta-Learning},
  author = {Zintgraf, Luisa and Shiarlis, Kyriacos and Kurin, Vitaly and Hofmann, Katja and Whiteson, Shimon},
  booktitle = {Proceedings of the 36th International Conference on Machine Learning (ICML)},
  series = {Proceedings of Machine Learning Research},
  volume = {97},
  pages = {7693--7702},
  year = {2019},
  publisher = {PMLR}
}

@article{youssry2024graybox,
  title     = {Experimental graybox quantum system identification and control},
  author    = {Youssry, Akram and Yang, Yang and Chapman, Robert J. and Haylock, Ben and Lenzini, Francesco and Lobino, Mirko and Peruzzo, Alberto},
  journal   = {npj Quantum Information},
  volume    = {10},
  pages     = {9},
  year      = {2024},
  publisher = {Springer Nature},
  doi       = {10.1038/s41534-023-00795-5}
}

@article{genois2025quantum,
  title = {Quantum optimal control of superconducting qubits based on machine-learning characterization},
  author = {Genois, \'Elie and Stevenson, Noah J. and Goss, Noah and Siddiqi, Irfan and Blais, Alexandre},
  journal = {Phys. Rev. Appl.},
  volume = {24},
  issue = {3},
  pages = {034073},
  numpages = {16},
  year = {2025},
  month = {Sep},
  publisher = {American Physical Society},
  doi = {10.1103/d9yg-d3qr},
  url = {https://link.aps.org/doi/10.1103/d9yg-d3qr}
}

@article{sivak2022modelfree,
    author = {Sivak, V. V and Eickbusch, A. and Liu, H. and Royer, B. and Tsioutsios, I. and Devoret, M. H},
    year = {2022},
    month = {03},
    pages = {},
    title = {Model-Free Quantum Control with Reinforcement Learning},
    volume = {12},
    journal = {Physical Review X},
    doi = {10.1103/PhysRevX.12.011059},
    url = {https://doi.org/10.1103/PhysRevX.12.011059}
}

@article{manohar2018sparse,
  author={Manohar, Krithika and Brunton, Bingni W. and Kutz, J. Nathan and Brunton, Steven L.},
  journal={IEEE Control Systems Magazine}, 
  title={Data-Driven Sparse Sensor Placement for Reconstruction: Demonstrating the Benefits of Exploiting Known Patterns}, 
  year={2018},
  volume={38},
  number={3},
  pages={63-86},
  doi={10.1109/MCS.2018.2810460}}

@article{eldar1997fps,
  title = {The farthest point strategy for progressive image sampling},
  author = {Eldar, Yuval and Lindenbaum, Michael and Porat, Moshe and Zeevi, Yehoshua Y.},
  journal = {IEEE Transactions on Image Processing},
  volume = {6},
  number = {9},
  pages = {1305--1315},
  year = {1997},
  publisher = {IEEE},
  doi = {10.1109/83.623193}
}

@article{mundada2019,
  title = {Suppression of Qubit Crosstalk in a Tunable Coupling Superconducting Circuit},
  author = {Mundada, Pranav and Zhang, Gengyan and Hazard, Thomas and Houck, Andrew},
  journal = {Phys. Rev. Appl.},
  volume = {12},
  issue = {5},
  pages = {054023},
  numpages = {10},
  year = {2019},
  month = {Nov},
  publisher = {American Physical Society},
  doi = {10.1103/PhysRevApplied.12.054023},
  url = {https://link.aps.org/doi/10.1103/PhysRevApplied.12.054023}
}

@article{lowdin1950non,
  title={On the non-orthogonality problem connected with the use of atomic wave functions in the theory of molecules and crystals},
  author={L{\"o}wdin, Per-Olov},
  journal={The Journal of Chemical Physics},
  volume={18},
  number={3},
  pages={365--375},
  year={1950},
  publisher={American Institute of Physics},
  doi={https://doi.org/10.1063/1.1747632},
  url={https://doi.org/10.1063/1.1747632}
}

@article{genois2021,
  title = {Quantum-Tailored Machine-Learning Characterization of a Superconducting Qubit},
  author = {Genois, \'Elie and Gross, Jonathan A. and Di Paolo, Agustin and Stevenson, Noah J. and Koolstra, Gerwin and Hashim, Akel and Siddiqi, Irfan and Blais, Alexandre},
  journal = {PRX Quantum},
  volume = {2},
  issue = {4},
  pages = {040355},
  numpages = {17},
  year = {2021},
  month = {Dec},
  publisher = {American Physical Society},
  doi = {10.1103/PRXQuantum.2.040355},
  url = {https://link.aps.org/doi/10.1103/PRXQuantum.2.040355}
}

@INPROCEEDINGS{Bayraktar2023,
  author={Bayraktar, Harun and Charara, Ali and Clark, David and Cohen, Saul and Costa, Timothy and Fang, Yao-Lung L. and Gao, Yang and Guan, Jack and Gunnels, John and Haidar, Azzam and Hehn, Andreas and Hohnerbach, Markus and Jones, Matthew and Lubowe, Tom and Lyakh, Dmitry and Morino, Shinya and Springer, Paul and Stanwyck, Sam and Terentyev, Igor and Varadhan, Satya and Wong, Jonathan and Yamaguchi, Takuma},
  booktitle={2023 IEEE International Conference on Quantum Computing and Engineering (QCE)}, 
  title={cuQuantum SDK: A High-Performance Library for Accelerating Quantum Science}, 
  year={2023},
  volume={01},
  number={},
  pages={1050-1061},
  doi={10.1109/QCE57702.2023.00119}}

\end{document}